\documentclass[prd,showpacs,showkeys,nofootinbib,twocolumn]{revtex4}

\begin{document}

\title{Multipolar equations of motion for extended test bodies in General Relativity}

\author{Jan Steinhoff}
\email{jan.steinhoff@uni-jena.de}
\homepage{http://www.tpi.uni-jena.de/gravity/People/steinhoff}
\affiliation{Theoretisch-Physikalisches Institut, Friedrich-Schiller-Universit\"at, Max-Wien-Platz 1, 07743 Jena, Germany}

\author{Dirk Puetzfeld}
\email{dirk.puetzfeld@aei.mpg.de}
\homepage{http://www.aei.mpg.de/~dpuetz/}
\affiliation{Max-Planck-Institute for Gravitational Physics (Albert-Einstein-Institute), Am Muehlenberg 1, 14476 Golm, Germany}

\date{ \today}

\begin{abstract}
We derive the equations of motion of an extended test body in the context of Einstein's theory of gravitation. The equations of motion are obtained via a multipolar approximation method and are given up to the quadrupolar order. Special emphasis is put on the explicit construction of the so-called canonical form of the energy-momentum density. The set of gravitational multipolar moments and the corresponding equations of motion allow for a systematic comparison to competing multipolar approximation schemes.
\end{abstract}


\maketitle


\section{Introduction}\label{introduction_sec}

The description of the motion of extended bodies in Einstein's theory of gravitation is a complicated and many-faceted problem. Nearly all applications of General Relativity crucially depend on our ability to describe how matter moves under the influence of the gravitational field. 

When it comes to the description of extended bodies, one usually has to resort to the use of approximation schemes due to the complexity of the theory. In this work we utilize a multipolar approximation method, originally devised by Tulczyjew \cite{Tulczyjew:1959}, to characterize the motion of extended test bodies. We explicitly work out the equations of motion at the monopolar, dipolar, as well as quadrupolar order with the help of this method.  In doing so, we put particular emphasis on the definition of multipole moments, as well as on the definition of combined quantities, e.g.\ the mass or the spin of objects, at different orders. The process of canonicalization is carried out in detail, and the resulting equations of motion are worked out in a fashion which allows for an easy comparison of quantities at different orders. 

Without going into historical detail we would like to point out, that multipolar methods were among the first methods to be studied in the context of the problem of motion in General Relativity. In particular Tulczyjew's method can be traced back to the seminal work of Mathisson \cite{Mathisson:1937}. We only note in passing, that several of the pioneering concepts of Mathisson's approach to the problem also resurface in other (later) works. For a more detailed account on the history of different multipolar approximation schemes we refer the reader to \cite{Puetzfeld:2009:1}.

Although we solely focus on the description of the motion of test bodies, the results obtained here are also relevant for other approximation methods. Most important are the post-Minkowskian \cite{Kerr:1959:1,Kerr:1959:2,Anderson:1976:2} and the post-Newtonian approximations -- see, e.g., the reviews \cite{AsadaFutamase1997,Blanchet:2006,Futamase:Itoh:2007} and references therein. The former method is useful to describe the scattering of an unbound and the latter method is commonly used to describe the inspiral of a bound system of two compact objects. For both approximation schemes the test mass results in the present paper can be extended to self-gravitating objects, if one relies on certain regularization techniques. Furthermore, our results can be used as input for higher order black hole perturbation schemes \cite{Mino:etal:1996:1,Tanaka:etal:1996:1,Mino:etal:1997:1} or other radiation reaction calculations \cite{Mino:etal:1997:2,Mino:etal:1997:3}. All of the above mentioned methods are used to model gravitational waves emitted from different astrophysical sources and are therefore of direct relevance for the field of gravitational wave astronomy \cite{Schutz:1999}.   

The structure of the paper is as follows. In section \ref{theorems_sec} we briefly recapitulate the basic ingredients of Tulczyjew's multipolar scheme. This is followed by the decompositions of a general set of moments in section \ref{sec_vel_decompositions}, which are crucial for the derivations in the succeeding sections. In sections \ref{pole_section} -- \ref{pole_dipole_quadrupole_section} we explicitly derive the monopolar, dipolar, as well as quadrupolar equations of motion. In \ref{comparison_sec} a detailed comparison to the multipolar approximation scheme by Dixon is performed. We draw our final conclusion in section \ref{conclusions_sec}. Appendices \ref{Jmoments}, \ref{combined_quant_app}, and \ref{dimension_acronyms_app} contain a brief overview of different quantities and our conventions as well as some useful transformation rules.   

\section{Basic definitions and theorems}\label{theorems_sec}

Conceptually the multipolar method of Tulczyjew \cite{Tulczyjew:1959} is based on the assumption, that the motion of an extended test body along a representative worldline can be characterized by a set of multipolar moments, which are built from the energy-momentum tensor $T^{ab}$ of the body. 

As in the case of other multipolar approximation schemes in the context of General Relativity, the starting point is the covariant conservation of the energy-momentum tensor, i.e.\
\begin{eqnarray}
\nabla_b T^{ab} = 0. \label{general_em_conservation}
\end{eqnarray}
The general strategy consists of working out the constraints of this equation on a general set of multipolar moments, which are covariantly defined via an expansion of the energy-momentum density of the following form: 
\begin{eqnarray}
\widetilde{T}^{ab} &=& \int_{-\infty}^{+\infty} \left\{ t^{ab} \delta_{(4)} + \nabla_c \left[ t^{cab} \delta_{(4)} \right] \right. \nonumber \\ 
&& \left.+ \nabla_d \nabla_c \left[ t^{dcab} \delta_{(4)} \right] + \dots \right\} ds. \label{general_decomp_em_density}
\end{eqnarray}
Here the $t^{abcd\dots}$ are general multipole moments, $\delta_{(4)} = \delta_{(4)}\left(x^a - Y^a \right)$ is the delta function, $Y^a\left(s\right)$ characterizes a representative worldline of the body, and $s$ denotes the proper time along this line. In other words, the continuous energy-momentum density $\widetilde{T}^{ab}$ of the extended body under consideration is replaced by an infinite set of moments, defined along a worldline -- which is completely arbitrary at the current stage. It is natural to expect, that the general energy-momentum conservation law (\ref{general_em_conservation}), imposes constraints on the moments defined via (\ref{general_decomp_em_density}), which in turn characterize certain properties of the body and its motion through spacetime.    

Of course, there is no simplification of the general problem of motion at the present stage. Insertion of the full representation (\ref{general_decomp_em_density}) into (\ref{general_em_conservation}) just yields a very complicated differential equation for the moments $t^{abcd\dots}$, which is actually of infinite order. The approximation comes from the fact, that one truncates the series in (\ref{general_decomp_em_density}), and only considers a finite number of moments. The general expectation is, that certain features of the body and its motion are adequately characterized by a small number of moments. This is of course analogous to multipolar expansion techniques as used in the context of classical mechanics or electrodynamics. In the following, the order of approximation is characterized by the notion of single-pole, dipole, or quadrupole, depending on which of the moments in the expansion (\ref{general_decomp_em_density}) are retained.     

On a technical level the method of Tulczyjew \cite{Tulczyjew:1959} is based on a generalization of the Du Bois-Reymond theorem -- called theorem B in the following, cf.\ section 3 in \cite{Tulczyjew:1959}. Before stating theorem B we need to define what Tulczyjew calls the {\it canonical form}. An arbitrary singular tensor density $\widetilde{A}^{c_1 \dots c_n}$ is said to be in canonical form if it can be written as 
\begin{eqnarray}
\widetilde{A}^{b_1 \dots b_n} = \sum_{k=0}^m \int_{-\infty}^{\infty} \nabla_{c_1 \dots c_k} \left[ \alpha^{c_1 \dots c_k b_1 \dots b_n} \delta_{(4)}\left(x^a - Y^{a} \right) \right], \nonumber \label{def_canonical_form}
\end{eqnarray}     
where the coefficients $\alpha^{c_1 \dots c_k b_1 \dots b_n}$ satisfy
\begin{eqnarray}
\alpha^{c_1 \dots c_k b_1 \dots b_n} = \alpha^{(c_1 \dots c_k) b_1 \dots b_n}, \label{canonical_form_cond_1} \\
u_{c_1} \alpha^{c_1 \dots c_k b_1 \dots b_n}=0. \label{canonical_form_cond_2}  
\end{eqnarray} 
Here we used the abbreviation $u^a:= d Y^{a} / d s$ for the tangent vector along the worldline. As was shown in \cite{Trautman:2002} it is always possible to achieve the canonical form.  

Theorem B states that, if for a tensor density $\widetilde{A}^{b_1 \dots b_n}$ and an arbitrary tensor field $T_{b_1 \dots b_n}$ we have  
\begin{eqnarray}
\int_D \widetilde{A}^{b_1 \dots b_n} T_{b_1 \dots b_n} = 0
\end{eqnarray}
in an arbitrary four-dimensional region $D$, then all the coefficients $\alpha^{c_1 \dots c_k b_1 \dots b_n}$ of the canonical form of the density $\widetilde{A}^{b_1 \dots b_n}$ vanish.

For further reading on the method of Tulczyjew, we refer readers to lecture V in \cite{Trautman:2002}.

\section{Decompositions with respect to the velocity}\label{sec_vel_decompositions}

Let us consider a set of general moments $t^{abc\dots}$ with the following symmetry properties:
\begin{eqnarray}
t^{c_1 \dots c_n ab} &=& t^{c_1 \dots c_n (ab)}, \nonumber \\
t^{c_1 \dots c_n ab} &=& t^{(c_1 \dots c_n) ab}. \label{sym_prop_moments}
\end{eqnarray}
Note that the symmetry in the second line is motivated by the integral representation of moments in the multipolar approximation scheme of Papapetrou \cite{Papapetrou:1951:3}, see also \cite{Obukhov:Puetzfeld:2009:1}. In the context of Tulczyjew's formalism there is a priori no reason to impose the symmetry in the first $n$ indices. One could carry out the calculation without imposing it. However, this would only lead to an unnecessary complication of the derivation, for the contributions from the antisymmetric parts of $t^{c_1 \dots c_n ab}$ could be absorbed in the course of the canonicalization procedure. We provide an explicit example of the absorption process in section \ref{pole_dipole_quadrupole_section}. 

With the help of the projector $\rho^a_b:=\delta^a_b - u^a u_b$ we may decompose the moments from (\ref{sym_prop_moments}) with respect to the four velocity, remember $u^a u_a = 1$, in the following way:
\begin{eqnarray}
t^{ab} &=&  \stackrel{0}{o}{\!}^{ab} + 2 \stackrel{0}{o}{\!}^{(a} u^{b)} + \stackrel{0}{t} u^a u^b, \nonumber \\
\stackrel{0}{o}{\!}^a    &:=& t^{cd} \rho^a_c u_d, \nonumber \\
\stackrel{0}{o}{\!}^{ab} &:=& t^{cd} \rho^a_c \rho^b_d,  \nonumber \\ 
\stackrel{0}{t}        &:=& t^{cd} u_c u_d. \label{vel_decomp_monopol} 
\end{eqnarray}
For the dipole moment, or three index quantity, we get
\begin{eqnarray}
t^{abc} &=& \stackrel{1}{o}{\!}^{abc} + 2 \stackrel{1}{o}{\!}^{a(b} u^{c)} + \stackrel{1}{o}{\!}^a u^b u^c + u^a \stackrel{1}{t}{\!}^{bc}, \nonumber \\
\stackrel{1}{o}{\!}^a       &:=& t^{def} \rho^a_d u_e u_f, \nonumber \\
\stackrel{1}{o}{\!}^{ab}    &:=& t^{def} \rho^a_d \rho^b_e u_f, \nonumber \\
\stackrel{1}{o}{\!}^{abc}   &:=& t^{def} \rho^a_d \rho^b_e \rho^c_f, \nonumber \\   
\stackrel{1}{t}{\!}^{bc}    &:=& t^{dbc} u_d. \label{vel_decomp_dipole}
\end{eqnarray}
For the quadrupole, or four index quantity, we get
\begin{eqnarray}
t^{abcd} &=& \stackrel{2}{o}{\!}^{abcd} + 2 \stackrel{2}{o}{\!}^{ab(c} u^{d)} + \stackrel{2}{o}{\!}^{ab} u^c u^d \nonumber \\
&& - u^a u^b \stackrel{2}{t}{\!}^{cd} + 2 u^{(a} \stackrel{2}{t}{\!}^{b)cd}, \nonumber \\
\stackrel{2}{o}{\!}^{ab}     &:=& t^{efgh} \rho^a_e \rho^b_f u_g u_h,\nonumber \\
\stackrel{2}{o}{\!}^{abc}    &:=& t^{efgh} \rho^a_e \rho^b_f \rho^c_g u_h,\nonumber \\
\stackrel{2}{o}{\!}^{abcd}   &:=& t^{efgh} \rho^a_e \rho^b_f \rho^c_g \rho^d_h,\nonumber \\
\stackrel{2}{t}{\!}^{cd}     &:=& t^{efcd} u_e u_f,\nonumber \\
\stackrel{2}{t}{\!}^{acd}    &:=& t^{eacd} u_e. \label{vel_decomp_quadrupole}
\end{eqnarray}
The decompositions in (\ref{vel_decomp_monopol})--(\ref{vel_decomp_quadrupole}) are going to play a central role in the upcoming derivations.

\section{Single-pole}\label{pole_section}

If we consider a single-pole object we start from
\begin{eqnarray}
0= \nabla_a \widetilde{T}^{ab} = \int \nabla_a \left[ t^{ab} \delta_{(4)} \right]. \label{pole_var_equation}
\end{eqnarray} 
In order to bring this equation into the canonical form, we decompose the integrand with respect to the first index, i.e.\ the one which is contracted with the derivative. Here and in the following we abbreviate the orthogonal projection of an index with respect to the velocity by a hat over the corresponding index, e.g., $t^{a \hat{b} c} := \rho^b_d t^{a d c}$. The decomposed version of (\ref{pole_var_equation}) then reads
\begin{eqnarray}
0= \int \nabla_a \left[ \left( t^{\hat{a}b} + u^a u_c t^{ c b} \right) \delta_{(4)} \right]. \label{decomp_pole_var_equation}
\end{eqnarray}
This integral can be split up by means of 
\begin{eqnarray}
\int \nabla_a \left[u^a T^{c_1 c_2 \dots } \delta_{(4)} \right] = \int \frac{\delta T^{c_1 c_2 \dots}}{ds} \delta_{(4)}, \label{lemma_c} 
\end{eqnarray}
as follows:
\begin{eqnarray}
0= \int \frac{\delta}{d s} \left[ u_c  t^{ c b} \right] \delta_{(4)} + \int \nabla_a \left[ t^{\hat{a}b} \delta_{(4)} \right]. \label{reduced_pole_var_equation}
\end{eqnarray}
This form of the integral allows for the application of theorem B, i.e.\ the equations of motion are now given by
\begin{eqnarray}
\frac{\delta}{d s} \left[ u_c  t^{ c b} \right] = 0, \quad \quad t^{\hat{a}b}=0. \label{pole_eom_after_tb_application} 
\end{eqnarray}
If we insert the orthogonal decomposition of $t^{ab}$ as given in (\ref{vel_decomp_monopol}) -- note that in the single-pole case we could have started right away with this decomposition, without making the intermediate step in (\ref{decomp_pole_var_equation}) -- the equations of motion in (\ref{pole_eom_after_tb_application}) take the form 
\begin{eqnarray}
\frac{\delta}{d s} \left[ \stackrel{0}{o}{\!}^{a} + u^a \stackrel{0}{t} \right]=0, \quad \quad \stackrel{0}{o}{\!}^{ab} + \stackrel{0}{o}{\!}^{a} u^b = 0. \label{pole_eom_after_complete_orthogonal_decomp}
\end{eqnarray}
From the second equation, due to the orthogonality, we can infer that
\begin{eqnarray}
\stackrel{0}{o}{\!}^{ab} = 0 \quad {\rm and} \quad \stackrel{0}{o}{\!}^{a} = 0, \label{conseq_mono_1st_order}
\end{eqnarray}
which leads to -- after reinsertion into the first equation in (\ref{pole_eom_after_complete_orthogonal_decomp})
\begin{eqnarray}
\stackrel{0}{t} = {\rm const} \quad {\rm and} \quad \frac{\delta}{d s} u^{a} = 0. \label{eom_pole_final}
\end{eqnarray}
In other words, we have shown that the equations of motion of a single-pole particle take the form of the geodesic equation. Equation (\ref{eom_pole_final}) suggests to identify the quantity $\stackrel{0}{t}$ with the mass $\stackrel{0}{m}$ of the test body. This result is of course not new, with the method outlined above, it was already derived by Tulczyjew in \cite{Tulczyjew:1959}. Finally, we note that the corresponding singular energy-momentum tensor is then given by
\begin{eqnarray}
\widetilde{T}^{ab} = \int  \stackrel{0}{m} u^a u^b \delta_{(4)}. \label{em_tensor_single_pole}
\end{eqnarray}

\section{Pole-dipole}\label{pole_dipole_section}

At the pole-dipole order we start from
\begin{eqnarray}
0= \nabla_a \widetilde{T}^{ab} = \int \nabla_a \left[ t^{ab} \delta_{(4)} \right] + \int \nabla_a \nabla_c \left[ t^{cab} \delta_{(4)} \right]. \label{dipole_var_equation}
\end{eqnarray} 
In order to be able to apply theorem B, we need to bring (\ref{dipole_var_equation}) into canonical form. First, we focus on the second term with the two covariant derivatives. Following the procedure outlined in the preceding section on the single-pole particle, we start with a decomposition of the indices which are contracted with the derivatives, i.e.\
\begin{eqnarray}
&& \int \nabla_a \nabla_c \left[ t^{cab} \delta_{(4)} \right] = \int \nabla_a \nabla_c \left[ \left( t^{\hat{c} \hat{a} b} + t^{\hat{c} d b} u^a u_d   \right) \delta_{(4)} \right] \nonumber \\
&& +  \int \nabla_a \nabla_c \left[ \left( t^{d \hat{a} b} u^c u_d + u^c u^a u_d u_e t^{d e b} \right) \delta_{(4)} \right] \nonumber \\
&& = \int \frac{\delta^2}{d s^2} \left(  t^{d e b} u_d u_e \right) \delta_{(4)} + \int \nabla_a \left[ \frac{\delta}{d s} \left( t^{d \hat{a} b} u_d \right) \delta_{(4)} \right. \nonumber \\ 
&& \left. + \frac{\delta u^a}{ds} u_d u_e t^{deb} \delta_{(4)} \right] + \int \nabla_a \nabla_c \left( t^{\hat{c} \hat{a} b} \delta_{(4)} \right) \nonumber \\
&&  + \int \nabla_a \nabla_c  \left( t^{\hat{c} d b} u^a u_d  \delta_{(4)} \right). \label{dipole_intermediate}
\end{eqnarray}
The last term in this equation can be rewritten with the help of
\begin{eqnarray}
\nabla_{[a} \nabla_{b]} T^{abc} = \frac{1}{2} R_{abd}{}^{c} T^{abd}, \label{antisym_cov_deriv}
\end{eqnarray}
and subsequent application of (\ref{lemma_c}) as follows: 
\begin{eqnarray}
&&\hspace{-0.6cm}\int \nabla_a \nabla_c \left[ t^{\hat{c}db} u^{a} u_d \right] \delta_{(4)} \nonumber \\
&&\hspace{-0.6cm}  = \int \nabla_c \left[\frac{\delta}{ds} \left(t^{\hat{c}db} u_d \right) \delta_{(4)} \right] + \int R_{ace}{}^{b} u^a u_d t^{\hat{c}de} \delta_{(4)}.  
\end{eqnarray} 
In order to arrive at the canonical form, one still needs to symmetrize the indices appearing in the second last term of (\ref{dipole_intermediate}). This again produces a curvature term at the lower order. Combining the rewritten form of (\ref{dipole_intermediate}) with the results at the single-pole order from the previous section, equation (\ref{dipole_var_equation}) takes the form
\begin{eqnarray}
&&\int \left[\frac{\delta^2}{ds^2} \left( t^{cdb} u_c u_d \right) + \frac{\delta}{ds} \left(t^{cb} u_c \right) \right. \nonumber\\
&&\left. \quad + \frac{1}{2} R_{ace}{}^b \left( 2 t^{\hat{c}de} u^a u_d + t^{\hat{c} \hat{a} e}\right) \right] \delta_{(4)} \nonumber\\
&&+ \int \nabla_a \left\{\left[\frac{\delta}{ds} \left( t^{d\hat{a}b} u_d + t^{\hat{a}db} u_d \right) + \frac{\delta u^a}{ds} u_d u_e t^{deb} \right. \right. \nonumber \\
&& \left. \left. + t^{\hat{a}b} \right] \delta_{(4)} \right\}  + \int \nabla_a \nabla_c \left[ t^{(\hat{c}\hat{a})b} \delta_{(4)} \right] = 0. \label{dipole_almost_canonical_form}
\end{eqnarray}
This is almost the canonical form, we still need to ensure the orthogonality of the first two terms in the second last line in (\ref{dipole_almost_canonical_form}). Once again we apply (\ref{lemma_c}) and obtain
\begin{eqnarray}
&&\int \nabla_a \left[\frac{\delta}{ds} \left( t^{d\hat{a}b} u_d + t^{\hat{a}db} u_d \right)  \delta_{(4)} \right] =   \nonumber \\
&&2 \int \nabla_a \left\{ \left[ \rho^a_c \frac{\delta}{ds} \left( t^{(cd)b} u_d \right) - \frac{\delta u^a}{ds} u_c u_d t^{(cd)b} \right] \delta_{(4)} \right\} \nonumber \\
&& -  2 \int \frac{\delta}{ds} \left(\frac{\delta u_c}{ds} u_d t^{(cd)b} \right) \delta_{(4)}.
\label{projector_pulled_out}
\end{eqnarray}
Reinsertion into (\ref{dipole_almost_canonical_form}) finally yields the canonical form of (\ref{dipole_var_equation}):
\begin{eqnarray}
&&\int \left[\frac{\delta^2}{ds^2} \left( t^{cdb} u_c u_d \right) + \frac{\delta}{ds} \left(t^{cb} u_c -2 \frac{\delta u_c}{ds} u_d t^{(cd)b}\right) \right. \nonumber\\
&&\left. \quad + \frac{1}{2} R_{ace}{}^b \left( 2 t^{\hat{c}de} u^a u_d + t^{\hat{c} \hat{a} e}\right) \right] \delta_{(4)} \nonumber\\
&&+ \int \nabla_a \left\{\left[ 2 \rho^a_c \frac{\delta}{ds} \left( t^{(cd)b} u_d \right) - \frac{\delta u^a}{ds} u_c u_d t^{dcb} \right. \right. \nonumber \\
&& \left. \left. + t^{\hat{a}b} \right] \delta_{(4)} \right\}  + \int \nabla_a \nabla_c \left[ t^{(\hat{c}\hat{a})b} \delta_{(4)} \right] = 0. \label{dipole_canonical_form}
\end{eqnarray}

With the help of theorem B -- starting at the highest order -- we can infer from (\ref{dipole_canonical_form}) that 
\begin{eqnarray}
0=t^{(\hat{c}\hat{a})b} = \stackrel{1}{o}{\!}^{(ca)b} + \stackrel{1}{o}{\!}^{(ca)} u^b. \label{dipole_conseq_1}
\end{eqnarray}
Here we made use of the decomposition (\ref{vel_decomp_dipole}) in the last step. Transvection of (\ref{dipole_conseq_1}) with the projector $\rho^d_b$ and the velocity $u_b$ yields two conditions, i.e.\
\begin{eqnarray}
\stackrel{1}{o}{\!}^{(ca)}=0,\quad {\rm and} \quad \stackrel{1}{o}{\!}^{(ca)b}=0. \label{dipole_conseq_1_1} 
\end{eqnarray}
From the last equation, together with the identity
\begin{eqnarray}
\stackrel{1}{o}{\!}^{cab} = \stackrel{1}{o}{\!}^{(ca)b} + \stackrel{1}{o}{\!}^{(bc)a} - \stackrel{1}{o}{\!}^{(ab)c}, \label{general_decomposition_o1}
\end{eqnarray}
we can infer that
\begin{eqnarray}
\stackrel{1}{o}{\!}^{cab} = 0. \label{dipole_conseq_1_2} 
\end{eqnarray}

At the second highest order theorem B yields
\begin{eqnarray}
2 \rho^a_c \frac{\delta}{ds} \left( t^{(cd)b} u_d \right) - \frac{\delta u^a}{ds} u_c u_d t^{dcb} + t^{\hat{a}b} =0. \label{dipole_conseq_2}
\end{eqnarray}
Insertion of the decompositions from (\ref{vel_decomp_monopol}) and (\ref{vel_decomp_dipole}) leads to
\begin{eqnarray}
\rho^a_c \frac{\delta}{ds} \left(\stackrel{1}{o}{\!}^{cb} + \stackrel{1}{o}{\!}^{c} u^b + \stackrel{1}{t}{\!}^{cb} \right) + \stackrel{0}{o}{\!}^{ab} + \stackrel{0}{o}{\!}^{a} u^b = 0. \label{dipole_conseq_2_1}
\end{eqnarray} 
Multiplication by $u_b$ and reinsertion of the result yields two equations. These allow us to express parts of the orthogonal decomposition of the single-pole moment in terms of the decompositions of the dipole moment as follows:  
\begin{eqnarray}
\stackrel{0}{o}{\!}^{a} &=& - u_d \rho^a_c \frac{\delta}{ds} \left(\stackrel{1}{o}{\!}^{cd} + \stackrel{1}{o}{\!}^{c} u^d + \stackrel{1}{t}{\!}^{cd} \right),  \label{dipole_conseq_2_2} \\
\stackrel{0}{o}{\!}^{ab} &=& - \rho^b_d \rho^a_c \frac{\delta}{ds} \left( \stackrel{1}{o}{\!}^{cd} + \stackrel{1}{o}{\!}^{c} u^d + \stackrel{1}{t}{\!}^{cd} \right).  \label{dipole_conseq_2_3}
\end{eqnarray}
Taking the antisymmetric part of (\ref{dipole_conseq_2_3}) yields
\begin{eqnarray}
\rho^b_d \rho^a_c \frac{\delta}{ds} \left( \stackrel{1}{o}{\!}^{[cd]} + \stackrel{1}{o}{\!}^{[c} u^{d]} \right) = 0. \label{dipole_conseq_2_4}
\end{eqnarray}
We introduce the {\it spin} in the following way:
\begin{eqnarray}
\stackrel{1}{S}{\!}^{ab}:= -2 \left(\stackrel{1}{o}{\!}^{[ab]} + \stackrel{1}{o}{\!}^{[a} u^{b]} \right). \label{def_spin}
\end{eqnarray}
Note that the prefactor is conventional, in particular the minus sign comes into play because we started with a positive sign in front of the dipole term in (\ref{dipole_var_equation}). Now (\ref{dipole_conseq_2_4}) turns into the well-known equation of motion for the spin \cite{Mathisson:1937,Papapetrou:1951:3}, i.e.\
\begin{eqnarray}
\rho^a_c \rho^b_d  \frac{\delta \stackrel{1}{S}{\!}^{cd}}{ds}  = \frac{\delta \stackrel{1}{S}{\!}^{ab}}{ds} - u^a u_c \frac{\delta \stackrel{1}{S}{\!}^{cb}}{ds} - u^b u_c \frac{\delta \stackrel{1}{S}{\!}^{ac}}{ds} = 0. \label{dipole_conseq_2_4_rewritten}
\end{eqnarray} 
Furthermore, if we make use of the first equation in (\ref{dipole_conseq_1_1}) and the definition of the spin (\ref{def_spin}), we can express parts of the orthogonal decomposition of the dipole moment in terms of the spin and the velocity as follows:
\begin{eqnarray}
\stackrel{1}{o}{\!}^{a} &=& - \stackrel{1}{S}{\!}^{ab} u_b, \label{o1a_in_terms_of_S}\\
\stackrel{1}{o}{\!}^{ab} &=& - \frac{1}{2} \stackrel{1}{S}{\!}^{ab} - u_c \stackrel{1}{S}{\!}^{c[a} u^{b]}. \label{o1ab_in_terms_of_S} 
\end{eqnarray}

From the lowest order in (\ref{dipole_canonical_form}) we get, again via theorem B and by insertion of the decompositions from (\ref{vel_decomp_monopol}) and (\ref{vel_decomp_dipole}), the following equation:\footnote{Here we introduced the shortcut ``$\dot{\phantom{T}}$''$:=\frac{\delta}{d s}$.}
\begin{eqnarray}
&&\hspace{-0.6cm}\frac{\delta}{ds} \left(u_d \frac{\delta}{ds} \stackrel{1}{t}{\!}^{db} + \stackrel{0}{o}{\!}^{b} + \stackrel{0}{t} u^{b} - \dot{u}_c \stackrel{1}{o}{\!}^{cb} - \dot{u}_c \stackrel{1}{o}{\!}^{c} u^b \right) \nonumber \\
&&\hspace{-0.6cm}+ \frac{1}{2} R_{ace}{}^b \left[2 u^a \left(\stackrel{1}{o}{\!}^{ce} + \stackrel{1}{o}{\!}^{c} u^e \right) + \stackrel{1}{o}{\!}^{cae} + \stackrel{1}{o}{\!}^{ca} u^e \right] = 0. \label{dipole_conseq_3_1}
\end{eqnarray}
Taking into account the symmetries of the quantities in this equation and our findings in (\ref{dipole_conseq_1_1}), (\ref{dipole_conseq_1_2}), (\ref{dipole_conseq_2_2}), (\ref{dipole_conseq_2_3}), (\ref{o1a_in_terms_of_S}), and (\ref{o1ab_in_terms_of_S}), we can rewrite (\ref{dipole_conseq_3_1}) as follows:
\begin{eqnarray}
\frac{\delta}{ds} \stackrel{1}{p}{\!}^b + \frac{1}{2} u^e \stackrel{1}{S}{\!}^{ac} R_{ace}{}^b =0. \label{dipole_conseq_3_2}
\end{eqnarray} 
This is the equation of motion for some kind of {\it generalized momentum}, which we define by
\begin{eqnarray}
\stackrel{1}{p}{\!}^{b}&:=& \left(\stackrel{0}{t} - u_c \dot{u}_d \stackrel{1}{S}{\!}^{cd} + u_c u_d \frac{\delta}{ds} \stackrel{1}{t}{\!}^{cd} \right)  u^b + u_d \frac{\delta}{ds} \stackrel{1}{S}{\!}^{bd} \nonumber \\
&=& \stackrel{1}{m} u^b + u_d \frac{\delta}{ds} \stackrel{1}{S}{\!}^{bd}. \label{def_generalized_momentum}
\end{eqnarray}
The second line serves as a definition of the {\it mass} $\stackrel{1}{m}$ which now -- in contrast to the result at the single-pole order -- contains also contributions from the spin as well as from the transversal component of the decomposition in (\ref{vel_decomp_dipole}).   

Our equations of motion for the pole-dipole test body in this section are the most general ones. {\it No} a priori restrictions were imposed on the decompositions in (\ref{vel_decomp_monopol}) and (\ref{vel_decomp_dipole}). Furthermore, it should be stressed that {\it no} assumptions were made regarding a possible spin supplementary condition. 

Equations (\ref{dipole_conseq_2_4_rewritten}) and (\ref{dipole_conseq_3_2}) are nowadays usually called the {\it Mathisson-Papapetrou} equations. In particular the characteristic spin-curvature coupling at the dipole-order is already present in Mathisson's pioneering work \cite{Mathisson:1937}. Note that Mathisson's equivalent to equation (\ref{dipole_conseq_3_2}) has a slightly different form. This is due to the fact that he sets the $\stackrel{1}{o}{\!}^{a}$ component in the orthogonal decomposition of the dipole moment to zero -- in other words he makes use of a supplementary condition -- at an early stage in his calculation. On the other hand, Papapetrou does not impose any supplementary condition in his derivation in \cite{Papapetrou:1951:3}. The equations of motion given by him are formally equivalent to (\ref{dipole_conseq_2_4_rewritten}) and (\ref{dipole_conseq_3_2}), but his moments are defined in a different way, cf.\ \cite{Puetzfeld:2009:1} for more details.

Finally, let us derive the energy-momentum tensor at the pole-dipole order. In terms of the spin, we have 
\begin{eqnarray}
\stackrel{1}{o}{\!}^{cd} + \stackrel{1}{o}{\!}^{c} u^d + \stackrel{1}{t}{\!}^{cd} = - \frac{1}{2} \stackrel{1}{S}{\!}^{cd} + u_e \stackrel{1}{S}{\!}^{e(c} u^{d)} + \stackrel{1}{t}{\!}^{cd}, \nonumber
\end{eqnarray}
hence (\ref{dipole_conseq_2_2}) and (\ref{dipole_conseq_2_3}) become
\begin{eqnarray}
\stackrel{0}{o}{\!}^{a} &=& u_d \rho^a_c \left(\frac{\delta}{ds} \stackrel{1}{S}{\!}^{cd} - \frac{\delta}{ds} {\stackrel{1}{t}}{}^{cd} \right) + \frac{1}{2} \dot{u}_d \rho^a_c \stackrel{1}{S}{\!}^{cd},\\
\stackrel{0}{o}{\!}^{ab} &=& - \rho^b_d \rho^a_c \frac{\delta}{ds} {\stackrel{1}{t}}{}^{cd}  + \rho^{(a}_c \dot{u}^{b)} \stackrel{1}{S}{\!}^{ce} u_e. 
\end{eqnarray}
If we use this result -- as well as all the constraints on the components of the decompositions in (\ref{vel_decomp_monopol}) and (\ref{vel_decomp_dipole}) obtained in this section -- in (\ref{dipole_var_equation}), with the help of (\ref{lemma_c}) the singular energy-momentum tensor for pole-dipole particles becomes 
\begin{eqnarray}
\widetilde{T}^{ab} =&& \int  u^{(a} \stackrel{1}{p}{\!}^{b)} \delta_{(4)} - \int  \nabla_c \left( \stackrel{1}{S}{\!}^{c(a} u^{b)} \delta_{(4)} \right). \label{em_tensor_pole_dipole}
\end{eqnarray}
Note that (\ref{em_tensor_pole_dipole}) is not in canonical form.

\subsection{Supplementary conditions and conserved quantities}\label{supp_cond_conserved_quant_sec}

The system of equations in (\ref{dipole_conseq_2_4_rewritten}) and (\ref{dipole_conseq_3_2}) is under-determined. This is evident from the appearance of the projectors in equation (\ref{dipole_conseq_2_4_rewritten}). Thus, {\it supplementary conditions}, or {\it constitutive relations}, involving the spin are needed to close the system. Before we discuss the impact of different conditions, we rewrite the equations of motion as follows:
\begin{eqnarray}
\stackrel{1}{m} \dot{u}^a &=& - \frac{1}{2} u^e \stackrel{1}{S}{\!}^{dc} R_{dce}{}^a - \rho^a_b \frac{\delta}{ds} \left( u_c \frac{\delta}{ds} \stackrel{1}{S}{\!}^{bc} \right), \label{eom_1_1_rewritten}\\
\frac{\delta}{ds}\stackrel{1}{m} &=& -\dot{u}_c \frac{\delta}{ds} \left(u_b \stackrel{1}{S}{\!}^{bc} \right), \label{eom_1_2_rewritten}\\
\frac{\delta}{ds} \stackrel{1}{S}{\!}^{ab} &=& 2 \stackrel{1}{p}{\!}^{[a} u^{b]}. \label{eom_2_rewritten}
\end{eqnarray}  
The first two equations are obtained from the orthogonal decomposition of (\ref{dipole_conseq_3_2}). 

There are basically two covariant supplementary conditions at the pole-dipole order which have been studied in the literature, i.e.\ 
\begin{eqnarray}
\stackrel{1}{S}{\!}^{ab} u_b = 0 \quad (\ast), \quad \quad \stackrel{1}{S}{\!}^{ab} \stackrel{1}{p}{\!}_b = 0. \quad (\ast \ast) \label{sup_conditions}
\end{eqnarray}
To our knowledge, the first condition can be traced back to an early work of Frenkel \cite{Frenkel:1926}, and the idea for the second condition appeared first in a work by Synge \cite{Synge:1935} in a special-relativistic context, see also \cite{Mathisson:1937,Papapetrou:1939,Pryce:1948,Corinaldesi:Papapetrou:1951,Pirani:1956,Tulczyjew:1959}. For both conditions there exist constant quantities, namely
\begin{eqnarray}
\frac{\delta}{ds} \stackrel{1}{m} &\stackrel{\ast}{=}& 0, \label{sup1_conseq_1} \\
\frac{\delta}{ds} {\stackrel{1}{\underline{m}}} := \frac{\delta}{ds} \sqrt{\stackrel{1}{p}{\!}_a \stackrel{1}{p}{\!}^a} & \stackrel{\ast \ast}{=}& 0, \label{sup2_conseq_1} \\
2 \frac{\delta}{ds} \left( \stackrel{1}{S} \right)^2 := \frac{\delta}{ds} {\stackrel{1}{S}{\!}_{ab} \stackrel{1}{S}{\!}^{ab}} & \stackrel{\phantom{\ast} \ast \,\, \vee \,\, \ast \ast}{=}& 0. \label{sup2_conseq_2}
\end{eqnarray}
Note that without the imposition of any supplementary condition the derivative of the alternative mass parameter ${\stackrel{1}{\underline{m}}}$ fulfills
\begin{eqnarray}
\stackrel{1}{m} {\stackrel{1}{\underline{m}}} \frac{\delta}{ds} {\stackrel{1}{\underline{m}}} = \frac{\delta \stackrel{1}{p}{\!}_a}{ds}  \stackrel{1}{p}{\!}_b \frac{\delta}{ds} \stackrel{1}{S}{\!}^{ab}.
\end{eqnarray}

Furthermore, in case the background spacetime allows for a Killing vector field $\varphi^a$, the quantity
\begin{eqnarray}
\frac{\delta}{ds} \stackrel{1}{E}:=\frac{\delta}{ds} \left( \stackrel{1}{p}{\!}^a \varphi_a + \frac{1}{2} \stackrel{1}{S}{\!}^{ab} \nabla_a \varphi_b \right) = 0, \label{E_def_and_conservation_dipole}
\end{eqnarray}
is conserved. For other (non-linear) conserved quantities at the pole-dipole order see \cite{Ruediger:1981,Ruediger:1983} and references therein.

\section{Pole-dipole-quadrupole}\label{pole_dipole_quadrupole_section}

At the pole-dipole-quadrupole order the variational equation takes the form
\begin{eqnarray}
0= \nabla_a \widetilde{T}^{ab} &=& \int \nabla_a \left[ t^{ab} \delta_{(4)} \right] + \int \nabla_a \nabla_c \left[ t^{cab} \delta_{(4)} \right] \nonumber \\ 
&&+ \int \nabla_a \nabla_d \nabla_c \left[ t^{dcab} \delta_{(4)} \right] . \label{quadrupole_var_equation}
\end{eqnarray}
In order to bring this equation to canonical form, we focus on the third term in (\ref{quadrupole_var_equation}) and proceed along the same lines as in the single-pole as well as in the pole-dipole case. With the help of the projector, the quadrupole moment can be decomposed as follows:
\begin{eqnarray}
t^{dcab} &=& t^{\hat{d}\hat{c}\hat{a}b}	+ u^a t^{\hat{d}\hat{c}eb} u_e + u^c u^d t^{gf\hat{a}b} u_f u_g \nonumber \\
&&	+ u^a u^c u^d t^{gfeb} u_e u_f u_g \nonumber \\ 
&&+ u^a ( u^d t^{f\hat{c}eb} + u^c t^{\hat{d}feb} ) u_e u_f \nonumber \\
&&+ ( u^d t^{e\hat{c}\hat{a}b} + u^c t^{\hat{d}e\hat{a}b} ) u_e. \label{quadrupole_decomp_first_three}
\end{eqnarray}
Due to their length, we provide the canonical form for the separate terms in (\ref{quadrupole_decomp_first_three}). As in the previous cases, the canonical form is achieved by repeated application of (\ref{lemma_c}), and the generalized version of (\ref{antisym_cov_deriv}) for multiple derivatives.\footnote{Note that one has to be careful when it comes to the usage of the ``$\dot{\phantom{T}}$'' notation in combination with the hat ``$\hat{\phantom{T}}$'' notation for projected indices.} 
\begin{widetext}
\begin{eqnarray}
\int \nabla_a \nabla_d \nabla_c [ t^{\hat{d}\hat{c}\hat{a}b} \delta_{(4)} ] &=& \int \nabla_a \nabla_d \nabla_c \left\{ t^{(\hat{d}\hat{c}\hat{a})b} \delta_{(4)} \right\}+\int \nabla_c \left\{R_{ade}{}^{b} t^{\hat{c}\hat{d}\hat{a}e} \delta_{(4)}+ \frac{1}{3} R_{ade}{}^{\hat{c}}t^{\hat{e}\hat{d}\hat{a}b} \delta_{(4)} \right\} \nonumber \\
&&+\int \left\{ \frac{1}{3} \frac{\delta}{d s} \left[R_{ade}{}^{c} u_c t^{\hat{e}\hat{d}\hat{a}b}\right] \delta_{(4)}+ \frac{2}{3} R_{dae}{}^{b}{}_{;c} t^{\hat{c}\hat{d}\hat{a}e} \delta_{(4)} \right\}, \label{canonical_quad_1}
\end{eqnarray}
\begin{eqnarray}
\int \nabla_a \nabla_d \nabla_c [ u^a t^{\hat{d}\hat{c}eb} u_e \delta_{(4)} ]&=&
\int \nabla_{(d} \nabla_{c)} \left\{ \rho^c_e \rho^d_f \frac{\delta}{d s} \left[ t^{\hat{f}\hat{e}ab} u_a \right] \delta_{(4)} \right\}	+ \int \nabla_d \bigg\{ 2 R_{acf}{}^{b} u^a t^{\hat{d}\hat{c}ef} u_e \delta_{(4)}	+ R_{acf}{}^{\hat{d}} u^a t^{\hat{f}\hat{c}eb} u_e \delta_{(4)} \nonumber \\
&&- 2 \rho^d_e \frac{\delta}{d s} \left[ \dot{u}_c t^{\hat{e}\hat{c}ab} u_a \right] \delta_{(4)} \bigg\}+ \int \bigg\{ \frac{\delta}{d s} \left[R_{acf}{}^{d} u^a u_d t^{\hat{f}\hat{c}eb} u_e 	\right] \delta_{(4)} - R_{acf}{}^{b}{}_{;d} u^a t^{\hat{c}\hat{d}ef} u_e \delta_{(4)} \nonumber \\
&&+ 2 \frac{\delta}{d s} \left[ \dot{u}_d \dot{u}_c	t^{\hat{d}\hat{c}ab} u_a \right] \delta_{(4)}	- R_{dce}{}^{b} u^d \dot{u}_f t^{\hat{f}\hat{c}ae} u_a \delta_{(4)} \bigg\},		\label{canonical_quad_2} 
\end{eqnarray}
\begin{eqnarray}
\int \nabla_a \nabla_d \nabla_c [ u^c u^d t^{gf\hat{a}b} u_f u_g \delta_{(4)} ] &=& \int \nabla_{(a} \nabla_{d)} \left\{ \dot{u}^d	t^{gf\hat{a}b} u_f u_g \delta_{(4)} \right\} + \int \nabla_a \bigg\{\rho^a_e \frac{\delta^2}{d s^2} \left[t^{gf\hat{e}b} u_f u_g \right] \delta_{(4)} \bigg\} \nonumber \\
&&+ \int \left\{ \frac{\delta}{d s} \left[ u_a \frac{\delta^2}{d s^2} \left(t^{gf\hat{a}b} u_f u_g \right) \right] \delta_{(4)} + \frac{1}{2} R_{ade}{}^{b} \dot{u}^d t^{gf\hat{a}e} u_f u_g \delta_{(4)} \right\}, \label{canonical_quad_3}
\end{eqnarray}
\begin{eqnarray}
\int \nabla_a \nabla_d \nabla_c [ u^a u^c u^d t^{gfeb} u_e u_f u_g \delta_{(4)} ] &=& \int \nabla_a \bigg\{ 3 \dot{u}^a \frac{\delta}{d s} \left[t^{gfeb} u_e u_f u_g \right]\delta_{(4)}	+ 2 \rho^a_h \ddot{u}^{h} t^{gfeb} u_e u_f u_g\delta_{(4)} \bigg\} \nonumber \\
&&+\int \bigg\{ \frac{\delta^3}{d s^3} \left[t^{gfeb} u_e u_f u_g \right] \delta_{(4)}	- 2 \frac{\delta}{d s} \left[ \dot{u}^a \dot{u}_a t^{gfeb} u_e u_f u_g \right] \delta_{(4)} \nonumber \\
&&+ R_{adc}{}^{b} \dot{u}^d u^a t^{gfec} u_e u_f u_g \delta_{(4)}\bigg\}, \label{canonical_quad_4}
\end{eqnarray}
\begin{eqnarray}
\int \nabla_a \nabla_d \nabla_c [ u^a ( u^d t^{f\hat{c}eb}+ u^c t^{\hat{d}feb} ) u_e u_f \delta_{(4)} ] &=& \int \nabla_{(c} \nabla_{a)} \left\{ 2 \dot{u}^a	t^{f\hat{c}eb} u_e u_f \delta_{(4)} \right\} \nonumber \\
&&+ \int \nabla_c \bigg\{ 2 \rho^c_g \frac{\delta^2}{d s^2} \left[t^{f\hat{g}eb} u_e u_f\right] \delta_{(4)} - R_{gda}{}^{\hat{c}} u^a u^d t^{f\hat{g}eb} u_e u_f 	\delta_{(4)} \bigg\} \nonumber \\
&&+ \int \bigg\{ 2 \frac{\delta}{d s} \left( R_{acg}{}^{b} u^a \right)	t^{f\hat{c}eg} u_e u_f \delta_{(4)}	+ 2 \frac{\delta}{d s} \left[ u_c \frac{\delta^2}{d s^2} \left(	t^{f\hat{c}eb} u_e u_f \right)\right] \delta_{(4)} \nonumber \\
&&+ 3 R_{acg}{}^{b} u^a \frac{\delta}{d s} \left[  t^{f\hat{c}eg} u_e u_f \right] \delta_{(4)} + R_{cdg}{}^{b}{}_{;a} u^a u^d t^{f\hat{c}eg} u_e u_f \delta_{(4)} \bigg\}, \label{canonical_quad_5}
\end{eqnarray} 
\begin{eqnarray}
\int \nabla_a \nabla_d \nabla_c [ ( u^d t^{e\hat{c}\hat{a}b} + u^c t^{\hat{d}e\hat{a}b} ) u_e \delta_{(4)} ] &=& \int \nabla_{(d} \nabla_{c)} \left\{ 2 \rho^c_e \rho^d_f \frac{\delta}{d s}\left[ t^{a\hat{e}\hat{f}b} u_a \right] \delta_{(4)} \right\} \nonumber \\
&&+ \int \nabla_d \bigg\{ R_{fce}{}^{\hat{d}} u^f t^{a\hat{c}\hat{e}b} u_a \delta_{(4)} 	+ R_{fce}{}^{b} u^f t^{a\hat{c}\hat{d}e} u_a \delta_{(4)} - 4 \rho^d_e \frac{\delta}{d s} \left[ \dot{u}_c	t^{a(\hat{c}\hat{e})b} u_a \right] \delta_{(4)} \bigg\} \nonumber \\
&&+ \int \bigg\{ \frac{\delta}{d s} \left[ R_{dce}{}^{a} u_a u^d	t^{g\hat{c}\hat{e}b} u_g \right] \delta_{(4)}	+ 4 \frac{\delta}{d s} \left[ \dot{u}_c \dot{u}_d t^{a(\hat{c}\hat{d})b} u_a \right] \delta_{(4)} \nonumber \\
&&+ R_{dce}{}^{b} \rho^c_f \rho^d_g \frac{\delta}{d s} \left[ t^{a\hat{f}\hat{g}e} u_a \right] \delta_{(4)}- 2 R_{dce}{}^{b} u^d \dot{u}_f t^{a\hat{c}\hat{f}e} u_a \delta_{(4)} \bigg\}. \label{canonical_quad_6}
\end{eqnarray}
\end{widetext}
Together with our result in (\ref{dipole_canonical_form}), equations (\ref{canonical_quad_1})--(\ref{canonical_quad_6}) yield the canonical form of (\ref{quadrupole_var_equation}) -- which we do not display here explicitly due to its length. As in the previous sections, the next step in the derivation of the equations of motion consists in the insertion of the orthogonal decomposition from (\ref{vel_decomp_monopol})--(\ref{vel_decomp_quadrupole}). To save us some work at the quadrupole order, we are {\it not} going to use the general decompositions in (\ref{vel_decomp_monopol})--(\ref{vel_decomp_quadrupole}) directly in the following, but rather transform them to a more compact form first. The new decompositions shall be given by
\begin{eqnarray}
\mathsf{t}^{ab} &=& \stackrel{0}{n}{\!}^{ab} + 2 \stackrel{0}{n}{\!}^{(a} u^{b)} + \stackrel{0}{n}{\!} u^a u^b, \label{alt_vel_decomp_monopol} \\
\mathsf{t}^{cab} &=& \stackrel{1}{n}{\!}^{cab} + 2 \stackrel{1}{n}{\!}^{c(a} u^{b)}	+ \stackrel{1}{n}{\!}^c u^a u^b, \label{alt_vel_decomp_dipole}\\
\mathsf{t}^{dcab} &=& \stackrel{2}{n}{\!}^{dcab} + 2 \stackrel{2}{n}{\!}^{dc(a} u^{b)}	+ \stackrel{2}{n}{\!}^{dc} u^a u^b. \label{alt_vel_decomp_quadrupole}
\end{eqnarray}
As becomes apparent from (\ref{alt_vel_decomp_dipole}) and (\ref{alt_vel_decomp_quadrupole}), the $\stackrel{1}{t}$ and $\stackrel{2}{t}$ terms in (\ref{vel_decomp_dipole}) and (\ref{vel_decomp_quadrupole}) have been absorbed in the new decomposition. The explicit transformation laws between the old and the new decomposition read 
\begin{widetext}
\begin{eqnarray}
\stackrel{0}{n}{\!} &=& \stackrel{0}{t}	+ u_e u_f \bigg[ \frac{\delta \stackrel{1}{t}{\!}^{ef}}{d s} - \frac{\delta^2 \stackrel{2}{t}{\!}^{ef}}{d s^2} + 2 \frac{\delta}{d s} \bigg( u_g \frac{\delta \stackrel{2}{t}{\!}^{gef}}{d s} \bigg) + 2 u^g \stackrel{2}{t}{\!}^{cd(e} R_{gcd}{}^{f)} \bigg], \label{trans_rule_1} \\
\stackrel{0}{n}{\!}^a &=& \stackrel{0}{o}{\!}^a	+ \rho^a_e u_f \bigg[ \frac{\delta \stackrel{1}{t}{\!}^{ef}}{d s}	- \frac{\delta^2 \stackrel{2}{t}{\!}^{ef}}{d s^2}	+ 2 \frac{\delta}{d s} \bigg( u_g \frac{\delta \stackrel{2}{t}{\!}^{gef}}{d s} \bigg)	+ 2 u^g \stackrel{2}{t}{\!}^{cd(e} R_{gcd}{}^{f)} \bigg], \label{trans_rule_2} \\
\stackrel{0}{n}{\!}^{ab} &=& \stackrel{0}{o}{\!}^{ab}	+ \rho^a_e \rho^b_f \bigg[ \frac{\delta \stackrel{1}{t}{\!}^{ef}}{d s} - \frac{\delta^2 \stackrel{2}{t}{\!}^{ef}}{d s^2} + 2 \frac{\delta}{d s} \bigg( u_g \frac{\delta \stackrel{2}{t}{\!}^{gef}}{d s} \bigg) + 2 u^g \stackrel{2}{t}{\!}^{cd(e} R_{gcd}{}^{f)} \bigg], \label{trans_rule_3} \\
\stackrel{1}{n}{\!}^c &=& \stackrel{1}{o}{\!}^c	+ \rho^c_e u_f u_g \bigg( - \dot{u}^e \stackrel{2}{t}{\!}^{fg} + 2 \frac{\delta \stackrel{2}{t}{\!}^{efg}}{d s} \bigg), \label{trans_rule_4} \\
\stackrel{1}{n}{\!}^{ca} &=& \stackrel{1}{o}{\!}^{ca}	+ \rho^c_e \rho^a_f u_g \bigg( - \dot{u}^e \stackrel{2}{t}{\!}^{fg}	+ 2 \frac{\delta \stackrel{2}{t}{\!}^{efg}}{d s} \bigg), \label{trans_rule_5} \\
\stackrel{1}{n}{\!}^{cab} &=& \stackrel{1}{o}{\!}^{cab}	+ \rho^c_e \rho^a_f \rho^b_g \bigg( - \dot{u}^e \stackrel{2}{t}{\!}^{fg} + 2 \frac{\delta \stackrel{2}{t}{\!}^{efg}}{d s} \bigg), \label{trans_rule_6} \\
\stackrel{2}{n}{\!}^{ab} &=& \stackrel{2}{o}{\!}^{ab}, \label{trans_rule_7} \\
\stackrel{2}{n}{\!}^{cab} &=& \stackrel{2}{o}{\!}^{cab}, \label{trans_rule_8} \\
\stackrel{2}{n}{\!}^{dcab} &=& \stackrel{2}{o}{\!}^{dcab}. \label{trans_rule_9} 
\end{eqnarray}
\end{widetext}
Hence, the energy-momentum tensor is given by
\begin{equation}
\widetilde{T}^{ab} = \int \mathsf{t}^{ab} \delta_{(4)}
	+ \int \nabla_c \left[ \mathsf{t}^{cab} \delta_{(4)} \right]
	+ \int \nabla_d \nabla_c \left[ \mathsf{t}^{dcab} \delta_{(4)} \right]. \label{em_tensor_alt_moments}
\end{equation}
From the form of the decompositions of the moments in (\ref{alt_vel_decomp_monopol})--(\ref{alt_vel_decomp_quadrupole}) and in (\ref{quadrupole_decomp_first_three}) it becomes apparent, that only a small fraction of $\stackrel{2}{n}{\!}^{abc\dots}$ terms contributes to the canonical form of $\nabla_b \widetilde{T}^{ab}$, when we make use of the re-defined $\mathsf{t}^{abc\dots}$ moments. In particular, only the integrals in (\ref{canonical_quad_1}) and (\ref{canonical_quad_2}) yield non-vanishing contributions.

The canonical form of the derivative of (\ref{em_tensor_alt_moments}) becomes
\begin{widetext}
\begin{eqnarray}
&& \int \nabla_a \nabla_d \nabla_c \left\{ \mathsf{t}^{(\hat{d}\hat{c}\hat{a})b} \delta_{(4)} \right\} + \int \nabla_{d} \nabla_{c} \left\{ \left[ \rho^{(c}_e \rho^{d)}_f \frac{\delta}{d s} \left[ \mathsf{t}^{\hat{f}\hat{e}ab} u_a \right] + \mathsf{t}^{(\hat{c}\hat{d})b} \right] \delta_{(4)}  \right\}  \nonumber \\
&& + \int \nabla_d \left\{ \left[ \rho^d_e \frac{\delta}{ds} \left( \mathsf{t}^{eab} u_a - 2 \dot{u}_c \mathsf{t}^{\hat{e}\hat{c}ab} u_a \right) + R_{acf}{}^{b} \left( \mathsf{t}^{\hat{d}\hat{c}\hat{a}f} + 2 u^a \mathsf{t}^{\hat{d}\hat{c}ef} u_e \right) +  R_{acf}{}^{\hat{d}} \left( \frac{1}{3} \mathsf{t}^{\hat{f}\hat{c}\hat{a}b} +u^a \mathsf{t}^{\hat{f}\hat{c}eb} u_e  \right) + \mathsf{t}^{\hat{d}b} \right]\delta_{(4)} \right\} \nonumber \\
&&+ \int \left\{ \left[ \frac{\delta}{d s} \left(R_{acf}{}^{d} u^a u_d \mathsf{t}^{\hat{f}\hat{c}eb} u_e + \frac{1}{3} R_{ade}{}^{c} u_c \mathsf{t}^{\hat{e}\hat{d}\hat{a}b} + 2 \dot{u}_d \dot{u}_c	\mathsf{t}^{\hat{d}\hat{c}ab} u_a + \mathsf{t}^{cb} u_c - \dot{u}_c u_d \mathsf{t}^{cdb} \right) \right. \right. \nonumber \\
&& \left. \left. + \frac{2}{3} R_{dae}{}^{b}{}_{;c} \mathsf{t}^{\hat{c}\hat{d}\hat{a}e} - R_{acf}{}^{b}{}_{;d} u^a \mathsf{t}^{\hat{c}\hat{d}ef} u_e - R_{dce}{}^{b} u^d \dot{u}_f \mathsf{t}^{\hat{f}\hat{c}ae} u_a + \frac{1}{2} R_{ace}{}^b \left( 2 \mathsf{t}^{\hat{c}de} u^a u_d + \mathsf{t}^{\hat{c} \hat{a} e}\right) \right] \delta_{(4)}  \right\} = 0. \label{canonical_form_final_quadrupole_order_new_moments}
\end{eqnarray}
\end{widetext}
Application of theorem B and insertion of the decomposition from (\ref{alt_vel_decomp_quadrupole}), yields at the highest order:
\begin{eqnarray} 
\stackrel{2}{n}{\!}^{(dca)b} + \stackrel{2}{n}{\!}^{(dca)} u^b = 0. \label{conseq_quad_3rd_order}
\end{eqnarray}
Orthogonal decomposition with respect to the open index $b$ leads to two symmetry relations, i.e.
\begin{eqnarray} 
\stackrel{2}{n}{\!}^{(dca)b} = 0 \quad \quad {\rm and} \quad \quad \stackrel{2}{n}{\!}^{(dca)} = 0. \label{quad_sym_relations_1}
\end{eqnarray}
The first relation in (\ref{quad_sym_relations_1}) allows us to infer that
\begin{eqnarray} 
\stackrel{2}{n}{\!}^{dcab} = \stackrel{2}{n}{\!}^{abdc}, \label{quad_sym_relations_1_1_and_2}
\end{eqnarray}
hence $\stackrel{2}{n}{\!}^{abdc}$ has symmetries similar to the ones of the curvature tensor\footnote{Using (\ref{quad_sym_relations_1}), and remembering $\stackrel{2}{n}{\!}^{dcab} = \stackrel{2}{n}{\!}^{(dc)(ad)}$, one can check that $\stackrel{2}{n}{\!}^{adcb} + \stackrel{2}{n}{\!}^{cadb} + \stackrel{2}{n}{\!}^{bdca} + \stackrel{2}{n}{\!}^{cbda} = - 2 \stackrel{2}{n}{\!}^{dcab}$ holds. On the other hand, one also has $\stackrel{2}{n}{\!}^{dabc} + \stackrel{2}{n}{\!}^{bdac} + \stackrel{2}{n}{\!}^{cabd} + \stackrel{2}{n}{\!}^{bcad} = - 2 \stackrel{2}{n}{\!}^{abdc}$. The left hand sides of these equations are identical, thus (\ref{quad_sym_relations_1_1_and_2}) holds.}. 

At the second highest order, again by application of theorem B to (\ref{canonical_form_final_quadrupole_order_new_moments}), and insertion of the decompositions from (\ref{alt_vel_decomp_dipole}) and (\ref{alt_vel_decomp_quadrupole}), we obtain
\begin{eqnarray} 
\rho^c_e \rho^d_f \frac{\delta}{ds} \left(\stackrel{2}{n}{\!}^{feb} + \stackrel{2}{n}{\!}^{fe} u^b \right) + \stackrel{1}{n}{\!}^{(cd)b} + \stackrel{1}{n}{\!}^{(cd)} u^b = 0. \label{quad_const_relations_1}
\end{eqnarray}  
The orthogonal split of (\ref{quad_const_relations_1}) yields two ``constraint  equations'' -- here and in the following we are going to use this name for equations, which allow us to express certain parts of moments in terms of parts of higher order moments, e.g.\  $\stackrel{1}{n}{\!}^{abc} = \stackrel{1}{n}{\!}^{abc} \left(\stackrel{2}{n}{\!}^{abc},\stackrel{2}{n}{\!}^{ab} \right)$ -- namely:
\begin{eqnarray}
 \stackrel{1}{n}{\!}^{(cd)b}  &=& - \rho^c_e \rho^d_f \rho^b_g \frac{\delta}{ds} \left(	\stackrel{2}{n}{\!}^{feg} + \stackrel{2}{n}{\!}^{fe} u^g \right), \label{quad_const_relations_1_1_a} \\
\stackrel{1}{n}{\!}^{(cd)} &=& - \rho^c_e \rho^d_f u_b \frac{\delta}{ds} \left(\stackrel{2}{n}{\!}^{feb} + \stackrel{2}{n}{\!}^{fe} u^b \right). \label{quad_const_relations_1_2_a} 
\end{eqnarray}
Equation (\ref{quad_const_relations_1_1_a}) can be used to rewrite $\stackrel{1}{n}{\!}^{abc}$ as follows:
\begin{eqnarray}
\stackrel{1}{n}{\!}^{dcb} &=& \stackrel{1}{n}{\!}^{(dc)b} + \stackrel{1}{n}{\!}^{(bd)c} - \stackrel{1}{n}{\!}^{(cb)d} \nonumber \\ 
&=& \rho^c_e \rho^d_f \rho^b_g \frac{\delta}{ds} \left( 2 \stackrel{2}{n}{\!}^{egf} - \stackrel{2}{n}{\!}^{fe} u^g - \stackrel{2}{n}{\!}^{gf} u^e + \stackrel{2}{n}{\!}^{eg} u^f \right) \nonumber \\
&=& 2 \frac{\delta}{ds} \stackrel{2}{n}{\!}^{cbd} + 2 \left( \stackrel{2}{n}{\!}^{abd} u^c + \stackrel{2}{n}{\!}^{cad} u^b + \stackrel{2}{n}{\!}^{cba} u^d \right) \dot{u}_a \nonumber \\
&& - \stackrel{2}{n}{\!}^{dc} \dot{u}^b - \stackrel{2}{n}{\!}^{bd} \dot{u}^c + \stackrel{2}{n}{\!}^{cb} \dot{u}^d.
\label{quad_const_relations_1_1_1}
\end{eqnarray}
Note that in the second step we made use of the second relation in (\ref{quad_sym_relations_1}), which yields an algebraic symmetry for $\stackrel{2}{n}{\!}^{abc}$, i.e.\
\begin{eqnarray}
\stackrel{2}{n}{\!}^{abc} + \stackrel{2}{n}{\!}^{bca} + \stackrel{2}{n}{\!}^{cab} = 0.  
\end{eqnarray}
An analogous relation holds for $\stackrel{2}{n}{\!}^{abcd}$. 

Equation (\ref{quad_const_relations_1_2_a}) can be simplified as follows:
\begin{eqnarray}
\stackrel{1}{n}{\!}^{(dc)} = \left( \stackrel{2}{n}{\!}^{dca} - 2 \stackrel{2}{n}{\!}^{a(d} u^{c)} \right) \dot{u}_a - \frac{\delta}{ds} \stackrel{2}{n}{\!}^{dc}. \label{quad_const_relations_1_2_simplified}
\end{eqnarray}

Let us now turn to the next order, i.e.\ the second line in (\ref{canonical_form_final_quadrupole_order_new_moments}). Applying the same procedure as before we obtain:
\begin{eqnarray}
\hspace{-0.3cm}&&\stackrel{0}{n}{\!}^{db} + \stackrel{0}{n}{\!}^{d} u^b + \rho^d_e \frac{\delta}{ds} \left[ \stackrel{1}{n}{\!}^{eb} + \stackrel{1}{n}{\!}^e u^b - 2 \left( \stackrel{2}{n}{\!}^{ecb} + \stackrel{2}{n}{\!}^{ec} u^b \right) \dot{u}_c \right] \nonumber \\
\hspace{-0.3cm}&&+ R_{acf}{}^g \rho^d_g \left[ \frac{1}{3} \left( \stackrel{2}{n}{\!}^{fcab} + \stackrel{2}{n}{\!}^{fca} u^b \right) + \left( \stackrel{2}{n}{\!}^{fcb} + \stackrel{2}{n}{\!}^{fc} u^b \right) u^a \right] \nonumber \\
\hspace{-0.3cm}&&+ R_{acf}{}^b \left[ \stackrel{2}{n}{\!}^{dcaf} + \stackrel{2}{n}{\!}^{dca} u^f + 2 \left( \stackrel{2}{n}{\!}^{dcf} + \stackrel{2}{n}{\!}^{dc} u^f \right) u^a \right] = 0. \nonumber \\ 
\hspace{-0.3cm}\label{quad_const_relations_1x}
\end{eqnarray}
The orthogonal split of (\ref{quad_const_relations_1x}) yields two equations
\begin{eqnarray}
\stackrel{0}{n}{\!}^{db} &=&  - \rho^d_e \rho^b_f \frac{\delta}{ds} \left[ \stackrel{1}{n}{\!}^{ef} + \stackrel{1}{n}{\!}^e u^f	- 2 ( \stackrel{2}{n}{\!}^{ecf} + \stackrel{2}{n}{\!}^{ec} u^f ) \dot{u}_c \right] \nonumber \\
	&& - R_{acf}{}^g \left[ \rho^b_g \stackrel{2}{n}{\!}^{dcaf} + \delta^b_g \stackrel{2}{n}{\!}^{dca} u^f + 2 \rho^b_g \stackrel{2}{n}{\!}^{dcf} u^a \right. \nonumber \\ 
	&& \left. + 2 \delta^b_g \stackrel{2}{n}{\!}^{dc} u^f u^a + \frac{1}{3} \stackrel{2}{n}{\!}^{fcab} \rho^d_g + \stackrel{2}{n}{\!}^{fcb} \rho^d_g u^a \right],  \label{quad_const_relations_1_1} \\
\stackrel{0}{n}{\!}^{d} &=&	- \rho^d_e u_b \frac{\delta}{ds} \left[ \stackrel{1}{n}{\!}^{eb} + \stackrel{1}{n}{\!}^e u^b - 2 \left( \stackrel{2}{n}{\!}^{ecb} + \stackrel{2}{n}{\!}^{ec} u^b \right) \dot{u}_c \right] \nonumber \\
&& - R_{acf}{}^{g} \left[u_g \stackrel{2}{n}{\!}^{dcaf} + 2 \stackrel{2}{n}{\!}^{dcf} u^a u_g + \frac{1}{3} \rho^d_g \stackrel{2}{n}{\!}^{fca} \right. \nonumber \\ 
&& \left.  + \rho^d_g \stackrel{2}{n}{\!}^{fc} u^a \right]. \label{quad_const_relations_1_3}
\end{eqnarray}
The antisymmetric part of (\ref{quad_const_relations_1_1}) can be viewed as the direct generalization of (\ref{dipole_conseq_2_4}), and suggests that the spin at the quadrupole order should be defined as
\begin{eqnarray}
\stackrel{2}{S}{\!}^{ab}:=-2 \left[ \stackrel{1}{n}{\!}^{[ab]} + \stackrel{1}{n}{\!}^{[a} u^{b]}	- 2 ( \stackrel{2}{n}{\!}^{c[ab]} + \stackrel{2}{n}{\!}^{c[a} u^{b]} ) \dot{u}_c \right], \label{def_spin_quadrupole}
\end{eqnarray}
yielding
\begin{eqnarray}
0 &=&  - \frac{1}{2} \rho^d_e \rho^b_f \frac{\delta}{ds} \stackrel{2}{S}{\!}^{ef} + R_{acf}{}^g \left[ \rho^{[b}_g \stackrel{2}{n}{\!}^{d]caf} + \delta^{[b}_g \stackrel{2}{n}{\!}^{d]ca} u^f  \right. \nonumber \\ 
	&& \left. + 2 \rho^{[b}_g \stackrel{2}{n}{\!}^{d]cf} u^a + 2 \delta^{[b}_g \stackrel{2}{n}{\!}^{d]c} u^f u^a + \frac{1}{3} \stackrel{2}{n}{\!}^{fca[b} \rho^{d]}_g \right. \nonumber \\ 
	&& \left.	+ \stackrel{2}{n}{\!}^{fc[b} \rho^{d]}_g u^a \right] \\
  &=& - \frac{1}{2} \rho^d_e \rho^b_f \frac{\delta}{ds} \stackrel{2}{S}{\!}^{ef}	- \rho^{[d}_e \rho^{b]}_g R_{acf}{}^g \left[ \frac{4}{3} \stackrel{2}{n}{\!}^{eacf} \right. \nonumber \\ 
  && \left. + 4 \stackrel{2}{n}{\!}^{eaf} u^c + 2 \stackrel{2}{n}{\!}^{ea} u^f u^c \right].  \label{quad_const_relations_1_4}
\end{eqnarray}
In contrast to our findings at the dipole order (\ref{dipole_conseq_2_4_rewritten}), this propagation equation for the spin contains -- as expected -- also contributions from the quadrupole moment, which couple to the curvature of spacetime. 

From (\ref{def_spin_quadrupole}) and (\ref{quad_const_relations_1_2_simplified}) we obtain
\begin{eqnarray}
 \stackrel{1}{n}{\!}^a &=& 2 \stackrel{2}{n}{\!}^{ab} \dot{u}_b - \stackrel{2}{S}{\!}^{ab} u_b, \label{n1_in_terms_of_S_n2} \\
 \stackrel{1}{n}{\!}^{ab} &=&  u^{[b} \stackrel{2}{S}{\!}^{a]c} u_c - \frac{\delta}{ds} \stackrel{2}{n}{\!}^{ab} - \frac{1}{2} \stackrel{2}{S}{\!}^{ab} \nonumber \\
 &&  - 2 \left(\stackrel{2}{n}{\!}^{cba} + \stackrel{2}{n}{\!}^{c(a} u^{b)} \right) \dot{u}_c,\label{n1_ab_in_terms_of_S_n2}
\end{eqnarray}
which can be viewed as the generalizations of (\ref{o1a_in_terms_of_S}) and (\ref{o1ab_in_terms_of_S}) to the quadrupole order.

What remains to be analyzed is the lowest order in (\ref{canonical_form_final_quadrupole_order_new_moments}). Application of theorem B at this order yields
\begin{eqnarray}
&&\frac{\delta}{ds} \left\{ R_{ace}{}^d \left[u^a u_d \left(\stackrel{2}{n}{\!}^{ecb} + \stackrel{2}{n}{\!}^{ec} u^b \right)  \right. \right.  \nonumber \\ 
&& \left. \left. + \frac{1}{3} u_d \left(\stackrel{2}{n}{\!}^{ecab} + \stackrel{2}{n}{\!}^{eca} u^b \right)  \right] + 2 \dot{u}_d \dot{u}_c \left(\stackrel{2}{n}{\!}^{dcb} + \stackrel{2}{n}{\!}^{dc} u^b \right) \right. \nonumber \\
&&\left. - \dot{u}_c \left(\stackrel{1}{n}{\!}^{cb} + \stackrel{1}{n}{\!}^{c} u^b \right) + \stackrel{0}{n}{\!}^b + \stackrel{0}{n} u^b \right\} \nonumber \\
&& + R_{ace}{}^b{}_{;d} \left[\frac{2}{3} \left(\stackrel{2}{n}{\!}^{dace} + \stackrel{2}{n}{\!}^{dac} u^e \right) - u^a \left(\stackrel{2}{n}{\!}^{cde} + \stackrel{2}{n}{\!}^{cd} u^e \right) \right] \nonumber \\
&& + R_{ace}{}^b \left[u^a \left(\stackrel{1}{n}{\!}^{ce} + \stackrel{1}{n}{\!}^{c} u^e \right) + \frac{1}{2} \left(\stackrel{1}{n}{\!}^{cae} + \stackrel{1}{n}{\!}^{ca} u^e \right) \right. \nonumber \\ 
&& \left. - \dot{u}_f u^a \left(\stackrel{2}{n}{\!}^{fce} + \stackrel{2}{n}{\!}^{fc} u^e \right) \right] = 0. \label{quad_const_relations_2} 
\end{eqnarray}
This equation is the analogue to (\ref{dipole_conseq_3_1}) found at the pole-dipole order. With the help of (\ref{quad_const_relations_1_1_1}), (\ref{quad_const_relations_1_3}), (\ref{n1_in_terms_of_S_n2}), and (\ref{n1_ab_in_terms_of_S_n2}) it can be rewritten in terms of $S^{ab}$ and the $\stackrel{2}{n}{\!}^{a\dots}$ from the decomposition of the quadrupole moment in (\ref{alt_vel_decomp_quadrupole}). In order to keep the equations at a manageable size, we introduce the auxiliary quantity
\begin{eqnarray} 
A^{ab} &:=& \stackrel{1}{n}{\!}^{ab} + \stackrel{1}{n}{\!}^a u^b - 2 \left( \stackrel{2}{n}{\!}^{cab} + \stackrel{2}{n}{\!}^{ca} u^b \right) \dot{u}_c \label{aux_a_definition} \\
&=&- \frac{1}{2} \stackrel{2}{S}{\!}^{ab} + u_c \stackrel{2}{S}{\!}^{c(a} u^{b)} - \frac{\delta}{ds} \stackrel{2}{n}{\!}^{ab} \nonumber \\
&& + 2 \left( \stackrel{2}{n}{\!}^{abc} - \stackrel{2}{n}{\!}^{c(a} u^{b)} \right) \dot{u}_c,
\end{eqnarray}
with the properties
\begin{eqnarray}
u_a A^{ab} &=& 0, \\
\dot{u}_a A^{ab} &=& - u_a \dot{A}^{ab}, \\
u_b A^{ab} &=& \stackrel{1}{n}{\!}^a - 2 \stackrel{2}{n}{\!}^{ca} \dot{u}_c = u_b \stackrel{2}{S}{\!}^{ba}, \\
A^{[ab]} &=& - \frac{1}{2} \stackrel{2}{S}{\!}^{ab}, \\
u_a u_b \frac{\delta }{ds}A^{ab} &=& \stackrel{2}{S}{\!}^{ab} \dot{u}_a u_b.
\end{eqnarray}
This allows us to rewrite (\ref{quad_const_relations_1_1}) and (\ref{quad_const_relations_1_3}) as follows:
\begin{eqnarray}
\stackrel{0}{n}{\!}^{db} &=& -\dot{A}^{(db)} + u^{(b} \dot{A}^{d)e} u_e + u_e \dot{A}^{e(b} u^{d)}  \nonumber \\
&&- \dot{A}^{ef} u_e u_f u^b u^d + \rho^{(d}_e \rho^{b)}_g R_{acf}{}^g  \nonumber \\
&& \times \left[ \frac{2}{3} \stackrel{2}{n}{\!}^{fcae} + 2 \stackrel{2}{n}{\!}^{eac} u^f + 2 \stackrel{2}{n}{\!}^{ea} u^f u^c \right] \\
&=& - \frac{\delta}{ds} A^{(db)} + u^d u^b u_e u_f \frac{\delta A^{ef}}{ds} + \rho^{(d}_e \rho^{b)}_g R_{acf}{}^g \nonumber \\
&& \times \left[ \frac{2}{3} \stackrel{2}{n}{\!}^{fcae} + 2 \stackrel{2}{n}{\!}^{eac} u^f + 2 \stackrel{2}{n}{\!}^{ea} u^f u^c \right] \nonumber \\
&& + 2 u^{(d} \rho_e^{b)} u_f \left[ \frac{\delta A^{ef}}{ds} - \frac{\delta A^{[ef]}}{ds} \right], \label{n0_ab_quad}\\
\stackrel{0}{n}{\!}^{d} &=& 	- \rho^d_e u_b \frac{\delta}{ds} A^{eb}	- R_{acf}{}^g \left[ \stackrel{2}{n}{\!}^{dcaf} u_g \right. \nonumber \\
&& \left.  + 2 \stackrel{2}{n}{\!}^{dcf} u^a u_g + \frac{1}{3} \stackrel{2}{n}{\!}^{fca} \rho^d_g + \stackrel{2}{n}{\!}^{fc} \rho^d_g u^a \right]. \label{n0_a_quad}	
\end{eqnarray}
Hence equation (\ref{quad_const_relations_2}) turns into
\begin{widetext}
\begin{eqnarray}
0 &=& \frac{\delta}{ds} \left[ \stackrel{0}{n} u^b + \stackrel{0}{n}{\!}^b - A^{cb} \dot{u}_c	+ R_{acd}{}^e u_e \left( \frac{1}{3} \stackrel{2}{n}{\!}^{dcab} + \frac{1}{3} \stackrel{2}{n}{\!}^{dca} u^b	+ \stackrel{2}{n}{\!}^{dcb} u^a + \stackrel{2}{n}{\!}^{dc} u^b u^a \right) \right] \nonumber \\
&&- R_{ace}{}^b{}_{;d} \left( \frac{2}{3} \stackrel{2}{n}{\!}^{dcae} + \frac{2}{3} \stackrel{2}{n}{\!}^{dca} u^e + \stackrel{2}{n}{\!}^{dce} u^a + \stackrel{2}{n}{\!}^{dc} u^e u^a \right) + \frac{1}{2} R_{ace}{}^b ( 2 \stackrel{1}{n}{\!}^{ce} u^a + 2 \stackrel{1}{n}{\!}^c u^e u^a + \stackrel{1}{n}{\!}^{cae} + \stackrel{1}{n}{\!}^{ca} u^e )	\nonumber \\ 
&& - R_{ace}{}^b u^a \dot{u}_f \left( \stackrel{2}{n}{\!}^{fce} + \stackrel{2}{n}{\!}^{fc} u^e \right) \\
&=& \frac{\delta}{ds} \bigg[ \stackrel{0}{n} u^b - \rho^b_e u_d \dot{A}^{ed} - A^{cb} \dot{u}_c	+ R_{ace}{}^d u_d \left( \frac{1}{3} \stackrel{2}{n}{\!}^{ecab} + \frac{1}{3} \stackrel{2}{n}{\!}^{eca} u^b + \stackrel{2}{n}{\!}^{ecb} u^a + \stackrel{2}{n}{\!}^{ec} u^b u^a \right) \nonumber \\
&& - R_{ace}{}^d \left( \stackrel{2}{n}{\!}^{bcae} u_d + 2 \stackrel{2}{n}{\!}^{bce} u^a u_d + \frac{1}{3} \stackrel{2}{n}{\!}^{eca} \rho^b_d + \stackrel{2}{n}{\!}^{ec} \rho^b_d u^a \right) \bigg] \nonumber \\
&& + \frac{1}{2} R_{ace}{}^b \bigg[ 2 u^a u^{[e} \stackrel{2}{S}{\!}^{c]f} u_f - u^a \stackrel{2}{S}{\!}^{ce} + u^e u^{a} \stackrel{2}{S}{\!}^{cf} u_f  - 4 u^a \dot{u}_d ( \stackrel{2}{n}{\!}^{dec} + \stackrel{2}{n}{\!}^{d(c} u^{e)} ) - 2 u^a \frac{\delta }{ds} \stackrel{2}{n}{\!}^{ce} - \frac{1}{2} u^e \stackrel{2}{S}{\!}^{ca}	 \nonumber \\
&& + 2 u^e u^a \left( u_d \stackrel{2}{S}{\!}^{dc} + 2 \dot{u}_d \stackrel{2}{n}{\!}^{dc} \right) + 2 \frac{\delta }{ds} \stackrel{2}{n}{\!}^{aec}  + 2 \left( \stackrel{2}{n}{\!}^{dec} u^a + \stackrel{2}{n}{\!}^{adc} u^e + \stackrel{2}{n}{\!}^{aed} u^c \right) \dot{u}_d - \stackrel{2}{n}{\!}^{ca} \dot{u}^e - \stackrel{2}{n}{\!}^{ec} \dot{u}^a + \stackrel{2}{n}{\!}^{ae} \dot{u}^c   \nonumber \\ 
&&   + 2 u^e \dot{u}_d \stackrel{2}{n}{\!}^{dca} - 2 \left( \stackrel{2}{n}{\!}^{fce} + \stackrel{2}{n}{\!}^{fc} u^e \right) u^a \dot{u}_f  \bigg]  -  R_{ace}{}^b{}_{;d} \bigg[ \frac{2}{3} \stackrel{2}{n}{\!}^{dcae} + \frac{2}{3} \stackrel{2}{n}{\!}^{dca} u^e + \stackrel{2}{n}{\!}^{dce} u^a + \stackrel{2}{n}{\!}^{dc} u^e u^a \bigg] \\
&=& \frac{\delta}{ds} \bigg[ \stackrel{0}{n} u^b	+ 2 A^{[bd]} \dot{u}_d - \frac{\delta ( A^{bd} u_d )}{ds} + u_e u_d u^b \frac{\delta A^{ed}}{ds} \nonumber \\
&& + R_{ace}{}^d \left( \frac{4}{3} \stackrel{2}{n}{\!}^{abce} u_d + \frac{1}{3} \stackrel{2}{n}{\!}^{eca} u^b u_d	+ 4 \stackrel{2}{n}{\!}^{bae} u^c u_d + \stackrel{2}{n}{\!}^{ec} u^a u^b u_d - \frac{1}{3} \stackrel{2}{n}{\!}^{eca} \rho^b_d - \stackrel{2}{n}{\!}^{ec} u^a \rho^b_d \right) \bigg] \nonumber \\
&&+ R_{ace}{}^b \bigg[ \frac{1}{2} u^e \stackrel{2}{S}{\!}^{ac} - \frac{\delta ( \stackrel{2}{n}{\!}^{eca} )}{ds} - \frac{\delta ( \stackrel{2}{n}{\!}^{ec} u^a )}{ds} \bigg]- R_{ace}{}^b{}_{;d} \bigg[ \frac{2}{3} \stackrel{2}{n}{\!}^{dcae} + \frac{2}{3} \stackrel{2}{n}{\!}^{dca} u^e	+ \stackrel{2}{n}{\!}^{dce} u^a + \stackrel{2}{n}{\!}^{dc} u^e u^a \bigg] \\
&=& \frac{\delta}{ds} \left[ \left( \stackrel{0}{n} + \stackrel{2}{S}{\!}^{ac} \dot{u}_a u_c - \frac{2}{3} R_{ace}{}^d \stackrel{2}{n}{\!}^{eca} u_d \right) u^b	  + R_{ace}{}^d \left( \frac{4}{3} \stackrel{2}{n}{\!}^{abce} u_d 	+ 4 \stackrel{2}{n}{\!}^{bae} u^c u_d	+ \frac{4}{3} \stackrel{2}{n}{\!}^{aec} \rho^b_d + 2 \stackrel{2}{n}{\!}^{ae} u^c \rho^b_d \right) \right. \nonumber \\
&& \left. + u_a \frac{\delta}{ds}\stackrel{2}{S}{\!}^{ba} \right] + \frac{1}{2} R_{ace}{}^b u^e \stackrel{2}{S}{\!}^{ac}	- R_{ace}{}^b{}_{;d} \bigg[ \frac{2}{3} \stackrel{2}{n}{\!}^{dcae}	+ \frac{4}{3} \stackrel{2}{n}{\!}^{dce} u^a + \frac{4}{3} \stackrel{2}{n}{\!}^{aed} u^c	+ \stackrel{2}{n}{\!}^{dc} u^e u^a + \stackrel{2}{n}{\!}^{ae} u^c u^d \bigg]. \label{eom_explicit_quad}
\end{eqnarray}
\end{widetext}
This concludes the derivation of the equations of motion at the quadrupole order. Equation (\ref{eom_explicit_quad}) replaces (\ref{dipole_conseq_3_1}) as the new center-of-mass equation of motion. 

Analogously to the pole-dipole order, we introduce combined quantities for the mass $\stackrel{2}{m}$ and the {\it generalized} momentum $\stackrel{2}{p}{\!}^a$ at the quadrupole order, as follows:
\begin{eqnarray}
\stackrel{2}{m} &:=&  \stackrel{0}{n} + \stackrel{2}{S}{\!}^{ac} \dot{u}_a u_c - \frac{2}{3} R_{ace}{}^d \stackrel{2}{n}{\!}^{eca} u_d,\label{mass_def_quadrupole} \\
\stackrel{2}{p}{\!}^b &:=& \stackrel{2}{m} u^b + u_a \frac{\delta}{ds} \stackrel{2}{S}{\!}^{ba} + R_{ace}{}^d \left( \frac{4}{3} \stackrel{2}{n}{\!}^{abce} u_d   \right. \nonumber \\ 
&& \left. + 4 \stackrel{2}{n}{\!}^{bae} u^c u_d + \frac{4}{3} \stackrel{2}{n}{\!}^{aec} \rho^b_d + 2 \stackrel{2}{n}{\!}^{ae} u^c \rho^b_d \right) . \label{momentum_def_quadrupole}
\end{eqnarray}
With these definitions the propagation equations for the spin (\ref{quad_const_relations_1_4}) and the center-of-mass (\ref{eom_explicit_quad}) take the form
\begin{widetext}
\begin{eqnarray}
\rho^a_c \rho^b_d \frac{\delta}{ds} \stackrel{2}{S}{\!}^{cd} &=& \rho^{[a}_g \rho^{b]}_e R_{dcf}{}^g \left[ \frac{8}{3} \stackrel{2}{n}{\!}^{edcf}+ 8 \stackrel{2}{n}{\!}^{edf} u^c + 4 \stackrel{2}{n}{\!}^{ed} u^f u^c \right], \label{eom1_explicit_quad} \\
\frac{\delta }{ds} \stackrel{2}{p}{\!}^b &=& - \frac{1}{2} R_{ace}{}^b u^e \stackrel{2}{S}{\!}^{ac} + \nabla_d R_{ace}{}^b \left[ \frac{2}{3} \stackrel{2}{n}{\!}^{dcae} + \frac{4}{3} \stackrel{2}{n}{\!}^{dce} u^a + \frac{4}{3} \stackrel{2}{n}{\!}^{aed} u^c + \stackrel{2}{n}{\!}^{dc} u^e u^a + \stackrel{2}{n}{\!}^{ae} u^c u^d \right]. \label{eom2_explicit_quad}
\end{eqnarray}
\end{widetext}
Our equations of motion for the pole-dipole-quadrupole test body in this section are the most general ones. It should be stressed that {\it no} assumptions were made regarding a possible spin supplementary condition. 

Before we introduce a new combined quantity for the quadrupole moment, we make contact with the multipole formalism of Dixon in section \ref{comparison_sec}. This will allow us to bring the equations of motion (\ref{eom1_explicit_quad}) and (\ref{eom2_explicit_quad}) into a very compact form. 

From the system of equations (\ref{eom1_explicit_quad}) and (\ref{eom2_explicit_quad}) it becomes clear, that the evolution of quadrupole components is not constrained -- in the sense that there is no dedicated propagation equation for the $\stackrel{2}{n}{\!}^{ab\dots}$. Equation (\ref{general_em_conservation}) seems to generate only equations of the constraint type at higher orders. Furthermore, one is already forced to introduce a supplementary condition at the pole-dipole approximation in order to obtain a closed system of equations. Hence, it is rather natural to expect, that additional supplementary conditions, now also involving the higher order moments $\stackrel{2}{n}{\!}^{ab\dots}$, are needed at the quadrupolar order. The choice of such conditions depends on the type of body under consideration. We are not going to touch upon the question of possible choices for such a condition in this work.   

\subsection{``Non''-conserved quantities}\label{non-conserved_quant_sec}

In this section we calculate the derivatives of the masses $\stackrel{2}{m}$ and $\stackrel{2}{\underline{m}}$, the spin length $\stackrel{2}{S}$, and a combined quantity $\stackrel{2}{E}$ at the quadrupole order. As definitions for the quantities $\stackrel{2}{S}$ and $\stackrel{2}{E}$, we use expressions which are completely analogous to the ones introduced at the pole-dipole order, see (\ref{sup2_conseq_2}) and (\ref{E_def_and_conservation_dipole}). 
\begin{widetext}
\begin{eqnarray}
\frac{\delta}{ds} \stackrel{2}{m} &=& \dot{u}_b \frac{\delta}{ds} \left(u_a \stackrel{2}{S}{\!}^{ab} \right) + u_b R_{ace}{}^b{}_{;d} \left(\frac{2}{3} \stackrel{2}{n}{\!}^{dcae} + 2 \stackrel{2}{n}{\!}^{aed} u^c + \stackrel{2}{n}{\!}^{ae} u^c u^d \right) + R_{ace}{}^d \left(\frac{4}{3} \stackrel{2}{n}{\!}^{abce} u_d \dot{u}_b   \right. \nonumber \\ 
&& \left. + 4 \stackrel{2}{n}{\!}^{bae} \dot{u}_b u_d u^c + \frac{4}{3} \stackrel{2}{n}{\!}^{aec} \dot{u}_d +2 \stackrel{2}{n}{\!}^{ae} \dot{u}_d u^c \right), \label{deriv_m2_quad}\\
\frac{\delta}{ds} \stackrel{2}{\underline{m}} &:=& \frac{\delta}{ds} \sqrt{\stackrel{2}{p}{\!}_a \stackrel{2}{p}{\!}^a} = \left(\stackrel{2}{m} \stackrel{2}{\underline{m}} \right)^{-1} \left[\frac{\delta}{ds} \left(\stackrel{2}{S}{\!}^{ab} \stackrel{2}{p}{\!}_b \right) \frac{\delta}{ds} \stackrel{2}{p}{\!}_a - \frac{2}{3} \stackrel{2}{\underline{m}}{\!}^2 \, u_a \nabla_b R_{cde}{}^a \left(\stackrel{2}{n}{\!}^{bcde} + 3 \stackrel{2}{n}{\!}^{deb} u^c  \right. \right. \nonumber \\
&& \left. \left. + \frac{3}{2} \stackrel{2}{n}{\!}^{de} u^b u^c \right) - \frac{8}{3} \stackrel{2}{p}{\!}_{[b} \frac{\delta}{ds} \stackrel{2}{p}{\!}_{a]} R_{cde}{}^a \left(\stackrel{2}{n}{\!}^{bcde} + \stackrel{2}{n}{\!}^{edc} u^b +3 \stackrel{2}{n}{\!}^{bce} u^d + \frac{3}{2} \stackrel{2}{n}{\!}^{bc} u^d u^e + \frac{3}{2} \stackrel{2}{n}{\!}^{de} u^b u^c  \right) \right], \label{deriv_m2__quad}\\
2 \frac{\delta}{ds} \left( \stackrel{2}{S} \right)^2 &:=& \frac{\delta}{ds} \left( \stackrel{2}{S}{\!}_{ab} \stackrel{2}{S}{\!}^{ab} \right) = 4 \stackrel{2}{S}{\!}_{ab} \stackrel{2}{p}{\!}^a u^b + \frac{16}{3} \stackrel{2}{S}{\!}_{ab} R_{cde}{}^a  \left( \stackrel{2}{n}{\!}^{bcde} + \stackrel{2}{n}{\!}^{edc} u^b + 3 \stackrel{2}{n}{\!}^{bce} u^d \nonumber \right. \\
&& \left. + \frac{3}{2} \stackrel{2}{n}{\!}^{bc} u^d u^e + \frac{3}{2} \stackrel{2}{n}{\!}^{de} u^b u^c \right), \label{deriv_s2_quad} \\
\frac{\delta}{ds} \stackrel{2}{E} &:=& \frac{\delta}{ds} \left( \stackrel{2}{p}{\!}^a \varphi_a + \frac{1}{2} \stackrel{2}{S}{\!}^{ab} \nabla_a \varphi_b \right) = - \frac{2}{3} \left(\varphi_a \nabla_b R_{cde}{}^a + 2 R_{cde}{}^a \nabla_b \varphi_a \right) \left(\stackrel{2}{n}{\!}^{bcde} + \stackrel{2}{n}{\!}^{edc} u^b + 3 \stackrel{2}{n}{\!}^{bce} u^d  \nonumber \right. \\
&& \left. + \frac{3}{2} \stackrel{2}{n}{\!}^{bc} u^d u^e + \frac{3}{2} \stackrel{2}{n}{\!}^{de} u^b u^c \right) \label{deriv_e2_quad} .
\end{eqnarray}
\end{widetext}
As becomes clear from (\ref{deriv_m2_quad})--(\ref{deriv_e2_quad}) these quantities are {\it no} longer conserved at the quadrupolar order. Note that this observation is independent of the choice of supplementary condition for the spin. A direct generalization of (\ref{sup_conditions}) -- in terms of the quantities at the quadrupole order -- only nullifies the first terms in (\ref{deriv_m2_quad})--(\ref{deriv_s2_quad}). The conservation of the quantities in (\ref{deriv_m2_quad})--(\ref{deriv_e2_quad}) depends on the details of the extended test body under consideration. Since we did not introduce any specific supplementary condition for the quadrupole components, the lack of conserved quantities at the current order of approximation is not unexpected. 

\subsection{Comparison to Dixon's scheme}\label{comparison_sec}

Our starting point are the equations of motion given by Dixon in (13.7) and (13.8) of \cite{Dixon:1974:1}, see also (168) and (169), as well as (171) and (172) in \cite{Dixon:1979}. According to Dixon the spin and the center-of-mass equations of motion -- up to the quadrupole order -- can be written in the form
\begin{eqnarray}
\frac{\delta}{ds} S^{ab}&=& 2 p^{[a} u^{b]} + \frac{4}{3} R_{cde}{}^{[a} I^{b]cde}, \label{Dspin} \\
\frac{\delta  }{ds} p_a &=&\frac{1}{2} R_{abcd} u^b S^{cd} - \frac{1}{6} \nabla_a R_{bcde} I^{bedc}, \label{Dcom}
\end{eqnarray} 
where $I^{abcd}$ has the following symmetries:
\begin{eqnarray}
I^{abcd} &=& I^{(ab)(cd)} = I^{cdab}, \label{Isym_0} \\
I^{(abc)d} &=& 0 \quad \Leftrightarrow \quad I^{abcd} + I^{bcad} + I^{cabd} = 0. \label{Isym}
\end{eqnarray}
Note that $I^{abcd}$ has no orthogonality properties with respect to $u^a$, except the ones that can be deduced from its symmetries. In Dixon's work the equations of motion in (\ref{Dspin}) and (\ref{Dcom}) are also often written in terms of a different set of moments, termed $J$ by him. To allow for a direct comparison, we provide the explicit transformation rules between the $I$- and $J$-moments in appendix \ref{Jmoments}. 

Orthogonal decomposition of (\ref{Dspin}) leads to
\begin{eqnarray}
\rho^a_c \rho^b_d \frac{\delta}{ds} S^{cd} &=& \frac{4}{3} \rho^{[a}_f \rho^{b]}_g R_{cde}{}^f I^{gcde}, \label{Dspin2} \\
\rho^a_b p^b &=& \dot{S}^{ab} u_b - \frac{4}{3} u_b R_{cde}{}^{[a} I^{b]cde}. \label{Dspin2_1}
\end{eqnarray}
With the definition $m:= u_a p^a$ equation (\ref{Dspin2_1}) turns into 
\begin{eqnarray}
p^a = m u^a + \dot{S}^{ab} u_b - \frac{4}{3} u_b R_{cde}{}^{[a} I^{b]cde}. \label{momentum_dixon}
\end{eqnarray}
Further, equation (\ref{Dcom}) can be written as
\begin{eqnarray}
\frac{\delta }{ds} p^a = - \frac{1}{2} R_{cdb}{}^a u^b S^{cd} - \frac{1}{3} \nabla_b R_{cde}{}^a I^{bcde}. \label{Dcom2}
\end{eqnarray}
Insertion of (\ref{momentum_dixon}) into (\ref{Dcom2}) and orthogonal decomposition of (\ref{Dcom2}) leads to separate equations of motion for $u^a$ and $m$. This is completely analogous to the pole-dipole case, c.f.\ equation (\ref{eom_1_1_rewritten}) and (\ref{eom_1_2_rewritten}).

The form of Dixon's equations of motion in (\ref{Dspin2}) and (\ref{Dcom2}) is the most suitable one for a comparison with our results from the previous section. We introduce the orthogonal decomposition of $I^{abcd}$, with its symmetries from (\ref{Isym_0}) and (\ref{Isym}) already implemented,  
\begin{eqnarray}
I^{dcab} &=& Q^{dcab}	+ 2 Q^{dc(a} u^{b)} + 2 Q^{ab(d} u^{c)}	+ Q^{dc} u^a u^b \nonumber \\
         && + Q^{ab} u^d u^c	- 2 u^{(d} Q^{c)(a} u^{b)}, \label{Idecomp}
\end{eqnarray}
where\footnote{Note that a similar decomposition has also been introduced in \cite{Ehlers:Rudolph:1977}. Therein quantities analogous to the $Q^{abcd}$, $Q^{abc}$, and $Q^{ab}$ are called stress-, flow-, and mass-quadrupole. In contrast to our decomposition with respect to $u^a$, Ehlers and Rudolph use the vector $p^a$ to perform the orthogonal decomposition. Furthermore, in \cite{Ehlers:Rudolph:1977} the decomposition was applied to the $J$-moments instead of the $I$-moments, see also (\ref{Jdecomp}).} 

\begin{eqnarray}
Q^{abcd} &=& Q^{(ab)(cd)} = Q^{cdab}, \\
Q^{(abc)d} &=& 0 \quad \Leftrightarrow \quad	Q^{abcd} + Q^{bcad} + Q^{cabd} = 0, \\
Q^{abc} &=& Q^{(ab)c}, \\
Q^{(abc)} &=& 0 \quad \Leftrightarrow \quad	Q^{abc} + Q^{bca} + Q^{cab} = 0, \\
Q^{ab} &=& Q^{(ab)}, \label{Q_sym}
\end{eqnarray}
and all $Q^{ab\dots}$ are orthogonal to the four velocity $u^a$. Notice that the symmetries of the $Q$-moments are identical to the symmetries of the $\stackrel{2}{n}$-moments, which suggests that they are proportional to each other. 

With the help of (\ref{Idecomp}) and the momentum in terms of the $Q$-moments, i.e.\ 
\begin{eqnarray}
p^b &=& m u^b + \dot{S}^{ba} u_a + R_{ace}{}^d \left( \frac{2}{3} Q^{abce} u_d + 2 Q^{bae} u^c u_d \right. \nonumber \\ 
&& \left. + \frac{2}{3} Q^{aec} \rho^b_d + Q^{ae} u^c \rho^b_d \right), \label{p_Dixon_in_terms_of_Q} 
\end{eqnarray}
the equations of motion (\ref{Dspin2}) and (\ref{Dcom2}) turn into
\begin{eqnarray}
\rho^a_c \rho^b_d \frac{\delta S^{cd}}{ds} &=& \rho^{[a}_g \rho^{b]}_e R_{dcf}{}^g	\left[ \frac{4}{3} Q^{edcf} + 4 Q^{edf} u^c \right. \nonumber \\
&& \left. + 2 Q^{ed} u^f u^c \right], \label{Dspin3} \\
\frac{\delta }{ds} p^b &=& - \frac{1}{2} R_{ace}{}^b u^e S^{ac} + \nabla_d R_{ace}{}^b \left[ \frac{1}{3} Q^{dcae} \right. \nonumber \\ 
&& \left. + \frac{2}{3} Q^{dce} u^a + \frac{2}{3} Q^{aed} u^c + \frac{1}{2} Q^{dc} u^a u^e \right. \nonumber \\
&& \left. + \frac{1}{2} Q^{ae} u^c u^d \right]. \label{Dcom3}
\end{eqnarray}
Comparison of (\ref{Dspin3}) to our equation of motion for the spin (\ref{eom1_explicit_quad}) shows that the $Q$- and the $\stackrel{2}{n}$-moments only differ by a factor of $2$, i.e.
\begin{eqnarray}
\hspace{-0.6cm}I^{dcab} &=& 2 \left( \stackrel{2}{n}{\!}^{dcab}	+ 2 \stackrel{2}{n}{\!}^{dc(a} u^{b)} + 2 \stackrel{2}{n}{\!}^{ab(d} u^{c)}	\right. \nonumber \\ 
\hspace{-0.6cm}&& \left. + \stackrel{2}{n}{\!}^{dc} u^a u^b + \stackrel{2}{n}{\!}^{ab} u^d u^c	- 2 u^{(d} \stackrel{2}{n}{\!}^{c)(a} u^{b)} \right), \label{dixon_like_definition_of_I}
\end{eqnarray}
as well as $S^{ab}=\stackrel{2}{S}{\!}^{ab}$. Furthermore, comparison of our equation of motion for the center-of-mass, i.e.\ equation (\ref{eom2_explicit_quad}), with (\ref{Dcom3}) yields $m=\stackrel{2}{m}$, and $p^a = \stackrel{2}{p}{\!}^a$.

\subsection{Energy-momentum tensor}\label{em_tensor_quad_subsec}

Our findings in the previous section allow for a compact representation of the energy-momentum tensor. In terms of Dixon's moments the decomposition in (\ref{alt_vel_decomp_monopol})--(\ref{alt_vel_decomp_quadrupole}) takes the form
\begin{widetext}
\begin{eqnarray}
\mathsf{t}^{dcab} &=& \frac{1}{2} I^{dcab} - 2 \stackrel{2}{n}{\!}^{ab(d} u^{c)}	- \stackrel{2}{n}{\!}^{ab} u^d u^c + 2 u^{(d} \stackrel{2}{n}{\!}^{c)(a} u^{b)}, \\
\mathsf{t}^{cab}  &=& - S^{c(a} u^{b)}	+ \frac{\delta}{ds} \left( 2 \stackrel{2}{n}{\!}^{abc} + \stackrel{2}{n}{\!}^{ab} u^c - 2 \stackrel{2}{n}{\!}^{c(a} u^{b)} \right) + A^{(ab)} u^c, \\
\mathsf{t}^{ab}   
&=& u^{(a} p^{b)} + R_{cde}{}^{(a} \left[\frac{2}{3} \stackrel{2}{n}{\!}^{b)cde}	+ \frac{2}{3} u^{b)} \stackrel{2}{n}{\!}^{edc}+ 2 \stackrel{2}{n}{\!}^{b)cd} u^e+ 2 \stackrel{2}{n}{\!}^{b)c} u^e u^d	\right] - \frac{\delta A^{(ab)}}{ds}.
\end{eqnarray}
\end{widetext}
Simplification by means of, e.g.,
\begin{eqnarray}
\hspace{-1cm} && \int \nabla_d \nabla_c \left[ \mathsf{t}^{dcab} \delta_{(4)} \right] =	\int \nabla_d \nabla_c \left[ \frac{1}{2} I^{dcab} \delta_{(4)} \right]	\nonumber \\ 
\hspace{-1cm} && - \int \nabla_c \left[ \frac{\delta}{ds} \left( 2 \stackrel{2}{n}{\!}^{abc} + \stackrel{2}{n}{\!}^{ab} u^c - 2 \stackrel{2}{n}{\!}^{c(a} u^{b)} \right)  \delta_{(4)} \right] \nonumber \\
\hspace{-1cm} &&- \int \left[  u^c R_{cde}{}^{(a} \left( 2 \stackrel{2}{n}{\!}^{b)ed} - u^{b)} \stackrel{2}{n}{\!}^{ed} - \stackrel{2}{n}{\!}^{b)d} u^e \right) \right] \delta_{(4)},
\end{eqnarray}
leads to
\begin{eqnarray}
\hspace{-1cm}&&\widetilde{T}^{ab} = \int \left( u^{(a} p^{b)} + \frac{1}{3} R_{cde}{}^{(a} I^{b)cde} \right) \delta_{(4)}	\nonumber \\ 
\hspace{-1cm}&&- \int \nabla_c \left( S^{c(a} u^{b)} \delta_{(4)} \right) + \frac{1}{2} \int \nabla_d \nabla_c \left( I^{dcab} \delta_{(4)} \right). \label{SETI}
\end{eqnarray}
Notice that this expression is not in canonical form, however, it seems to be the simplest representation of the energy-momentum tensor density expressed in terms of the Dixon-moments. All constraint-type relations have been implemented into the energy-momentum tensor. Thus, all consequences of (\ref{general_em_conservation}) can equivalently be written in a compact way as the equations of motion (\ref{Dspin}) and (\ref{Dcom}), the symmetries of the quadrupole (\ref{Isym_0}) and (\ref{Isym}), and the energy-momentum tensor in the form (\ref{SETI}).

\section{Conclusions}\label{conclusions_sec}

In this work we explicitly derived the equations of motion for extended test bodies in General Relativity with the help of Tulczyjew's multipolar approximation method up to the quadrupolar order. To our knowledge this is the first time the method by Tulczyjew has been used beyond the pole-dipole order. In our derivation we put special emphasis on a transparent notation, which allows for a direct identification of the contributions of the multipole moments at different orders, and explicitly carried out the canonicalization process. We would like to stress again, that we did not make any assumption for a supplementary condition in our derivation of the equations of motion.

Our results are of direct relevance for other perturbation methods in the context of the general relativistic problem of motion. In particular the equations of motion in (\ref{eom1_explicit_quad}) and (\ref{eom2_explicit_quad}), or -- alternatively -- the ones given in (\ref{Dspin}) and (\ref{Dcom}), as well as the energy-momentum tensor in (\ref{SETI}), are needed in approximation schemes which aim for a description of self-gravitating compact objects and the gravitational radiation emitted by these systems.  

\subsection{Structure of the equations of motion}\label{structure_sec}

Table \ref{tab_structure} gives a compact overview of the different equations, which were obtained from the conservation law (\ref{general_em_conservation}) and the decomposition (\ref{general_decomp_em_density}) via repeated application of theorem B. We introduced the classification of two different types of equations, termed ``constraint'' and ``evolution''. As becomes clear from table \ref{tab_structure}, the same type of pattern of equations repeats at each multipolar order. In particular, one does {\it not} expect more than two equations of the evolution-type in the context of Tulczyjew's approximation scheme. This can be viewed as support of Dixon's result \cite{Dixon:1970:1,Dixon:1970:2,Dixon:1974:1,Dixon:1979}, who provides a solution of what he calls the {\it variational equations of mechanics}\footnote{This notion goes back to Mathisson \cite{Mathisson:1937}, see \cite{Puetzfeld:2009:1} for more historical details.} -- i.e.\ the combination of equation (\ref{general_em_conservation}) and (\ref{general_decomp_em_density}) -- to {\it any} order.

Our results in (\ref{eom1_explicit_quad}) and (\ref{eom2_explicit_quad}) also allow for a direct comparison to the quadrupolar equations of motion derived in \cite{Obukhov:Puetzfeld:2009:1}. The results therein were obtained via a different multipolar approximation scheme which goes back to Papapetrou \cite{Papapetrou:1951:3}. Although the same information is encoded in the system of equations -- Tulczyjew's as well as Papapetrou's method share the same starting point, i.e.\ the ``conservation'' of energy in the form of (\ref{general_em_conservation}), furthermore in both methods the full moments are taken into account -- the final representation of the equations of motion is different. This difference can mainly be ascribed to the use of the orthogonal decomposition of the moments, which is an integral part in Tuczyjew's procedure, and is introduced at a very early stage to support the derivation of the canonical form. In particular the recursive transfer of higher order moments to lower differential orders in the canonical form, see the final result (\ref{canonical_form_final_quadrupole_order_new_moments}) at the quadrupolar order in this respect, yields structurally different equations. While it is mainly a question of practicability which system of equations of motion should be given preference, the main benefit of the method by Tulczyjew is its intrinsic covariance as well as its systematic -- albeit somewhat laborious -- way to generate a hierarchical set of equations.  

\begin{table}
\caption{\label{tab_structure}Structure of the equations of motion.} 
\begin{ruledtabular}
\begin{tabular}{clcc}
Order&Quantity&Type&Equation\\
\hline
\multicolumn{4}{l}{{Single-pole}}\\ 
\hline
1&$\stackrel{0}{o}{\!}^{ab}$&C&(\ref{conseq_mono_1st_order})\\
1&$\stackrel{0}{o}{\!}^{a}$&C&(\ref{conseq_mono_1st_order})\\
0&$\stackrel{0}{t}{\left(\stackrel{0}{o}{\!}^{a}\right)}$&E&(\ref{pole_eom_after_complete_orthogonal_decomp})\\
&&&\\
\hline
\multicolumn{4}{l}{{Dipole}}\\ 
\hline
2&$\stackrel{1}{o}{\!}^{(ab)c}$&C&(\ref{dipole_conseq_1_1})\\
2&$\stackrel{1}{o}{\!}^{(ab)}$&C&(\ref{dipole_conseq_1_1})\\
1&$\stackrel{0}{o}{\!}^{ab}{\left(\stackrel{1}{o}{\!}^{ab},\stackrel{1}{o}{\!}^{a},\stackrel{1}{t}{\!}^{ab}\right)}$&E+C &(\ref{dipole_conseq_2_3})\\
1&$\stackrel{0}{o}{\!}^{a}{\left(\stackrel{1}{o}{\!}^{ab},\stackrel{1}{o}{\!}^{a},\stackrel{1}{t}{\!}^{ab}\right)}$&C&(\ref{dipole_conseq_2_2})\\
0&$\stackrel{0}{t}{\left(\stackrel{1}{o}{\!}^{ab},\stackrel{1}{o}{\!}^{a},\stackrel{1}{t}{\!}^{ab},\stackrel{0}{o}{\!}^{a}\right)}$&E&(\ref{dipole_conseq_3_1})\\
&&&\\
\hline
\multicolumn{4}{l}{{Quadrupole}}\\ 
\hline
3&$\stackrel{2}{n}{\!}^{(abc)d}$&C&(\ref{conseq_quad_3rd_order})\\
3&$\stackrel{2}{n}{\!}^{(abc)}$&C&(\ref{conseq_quad_3rd_order})\\
2&$\stackrel{1}{n}{\!}^{(ab)c}{\left(\stackrel{2}{n}{\!}^{abc},\stackrel{2}{n}{\!}^{ab}\right)}$&C&(\ref{quad_const_relations_1_1_a})\\
2&$\stackrel{1}{n}{\!}^{(ab)}{\left(\stackrel{2}{n}{\!}^{abc},\stackrel{2}{n}{\!}^{ab}\right)}$&C&(\ref{quad_const_relations_1_2_a})\\
1&$\stackrel{0}{n}{\!}^{ab}{\left(\stackrel{2}{n}{\!}^{abcd},\stackrel{2}{n}{\!}^{abc},\stackrel{2}{n}{\!}^{ab},\stackrel{1}{n}{\!}^{ab},\stackrel{1}{n}{\!}^{a}\right)}$&E+C&(\ref{quad_const_relations_1_1})\\
1&$\stackrel{0}{n}{\!}^{a}{\left(\stackrel{2}{n}{\!}^{abcd},\stackrel{2}{n}{\!}^{abc},\stackrel{2}{n}{\!}^{ab},\stackrel{1}{n}{\!}^{ab},\stackrel{1}{n}{\!}^{a}\right)}$&C&(\ref{quad_const_relations_1_3})\\
0&$\stackrel{0}{n}{\left(\stackrel{2}{n}{\!}^{abcd},\stackrel{2}{n}{\!}^{abc},\stackrel{2}{n}{\!}^{ab},\stackrel{1}{n}{\!}^{abc},\stackrel{1}{n}{\!}^{ab},\stackrel{1}{n}{\!}^{a}\right)}$&E&(\ref{quad_const_relations_2})\\
&&&\\
\hline
\multicolumn{4}{l}{{Type: ``C''=constraint, ``E''=evolution}}
\end{tabular}
\end{ruledtabular}
\end{table}

\subsection{Open problems}\label{open_problems_sec}

Albeit the results obtained in this work are complete, in the sense that they cover the quadrupolar order in the context of Tulczyjew's approximation scheme in the most general way, there remain several interesting open questions to be addressed. Some of these open problems are not specific to Tulczyjew's scheme, and are also inherent to other multipolar approximation schemes.

\subsection{Supplementary conditions}\label{open_problems_subsubsec_supp}

As we have seen in sections \ref{pole_dipole_section} and \ref{pole_dipole_quadrupole_section}, additional conditions are needed, starting at the pole-dipole order, to close the systems of the equations of motion. The necessity for a spin supplementary condition can be seen as connected to the fixation of a specific representative worldline inside the object. This is well-known in the special-relativistic context, where a spin supplementary condition selects a representative worldline and vice-versa, see, e.g., \cite{Moller:1949}. In General Relativity, however, things are more subtle, c.f.\ \cite{Beiglboeck:1965,Beiglboeck:1967,Ehlers:Rudolph:1977,Schattner:1978,Schattner:1979,Schattner:1979:1,Dixon:2008:1} in this respect. In particular, it is difficult to prove that a spin supplementary condition fixes a representative worldline in a {\it unique} way. This is of course directly related to the fact, that the quantities ${\stackrel{2}{p}{\!}^a}$ and ${\stackrel{2}{S}{\!}^{ab}}$ are combinations of geometrical quantities as well as of multipolar moments from different orders. Further studies are needed in the context of Tulczyjew's scheme when it comes to the choice of suitable supplementary conditions at higher orders. 

However, from a more pragmatic point of view every supplementary condition, which leads to a closed system of equations, can be used. The important question for applications is, whether the equations describe the motion in an accurate way.

The results derived in this paper could be interesting when studying whether two sets of moments contained in the energy-momentum tensor on the same, or infinitesimally close worldlines, are equivalent. In the latter case one first has to shift the distributional energy-momentum tensor from one of the worldlines to the other. If two sets of moments are equivalent, then the coefficients in the canonical form of the corresponding energy-momentum tensors must be the same (by virtue of theorem B). We will not work this out in detail here. However, an immediate consequence is that an infinitesimal change of the worldline directly translates into an infinitesimal change of the moment ${\stackrel{1}{n}{\!}^a}$. This has been used in \cite{Mathisson:1937,Tulczyjew:1959,Trautman:2002} to argue that, by a suitable choice of the representative wordline, one can restrict to ${\stackrel{1}{n}{\!}^a(\tau) = 0}$ for all $\tau$. In consideration of (\ref{n1_in_terms_of_S_n2}) a change of ${\stackrel{1}{n}{\!}^a}$ corresponds to a change of ${\stackrel{2}{S}{\!}^{ab} u_b}$ and will thus have an impact on the supplementary condition fulfilled by the spin. In this way, an infinitesimal change of the representative worldline can be related to an infinitesimal change of the spin supplementary condition.

\subsection{Combined and conserved quantities}\label{open_problems_subsubsec_cons}

While it is possible to generate a hierarchical set of equations of motion, c.f.\ table \ref{tab_structure}, with the help of Tulczyjew's method, it is non-trivial to devise ``combined'' quantities at higher multipolar orders. A good example is the quantity $I^{abcd}$ as introduced in equation (\ref{dixon_like_definition_of_I}). There is no straightforward algorithm in Tulczyjew's scheme which tells us how to construct it. While one may consider the introduction of quantities like $I^{abcd}$ as a mere question of taste -- after all the main benefit is a more compact form of the equations of motion -- it appears to be desirable to have such quantities at one's disposal. In particular when it comes to the search for conserved objects.   

As we have shown in section \ref{non-conserved_quant_sec}, several of the combined quantities are no longer conserved at the quadrupolar order. This is of course directly linked to the fact that we did not assume any supplementary condition in the course of our derivation. Even the direct generalization and use of the supplementary conditions from the pole-dipole order does not yield a set of conserved quantities at the quadrupolar order. Our results in (\ref{deriv_m2_quad})--(\ref{deriv_e2_quad}) make clear, that additional conditions for the quadrupole moments and/or symmetries of the underlying spacetime are needed to obtain a set of conserved quantities. 

\subsection{Further approximations}\label{open_problems_subsubsec_app}
 
Due to the complexity of the expressions at the quadrupole order, further approximations are needed for applications. Most interesting is the restriction to some kind of mass-quadrupole, i.e., neglecting the flow- and stress-quadrupole\footnote{Notice that there can be slightly different ways to define such mass-, flow-, and stress-quadrupoles. They only have to approach the correct Newtonian limit.}. Corrections from the mass-quadrupole are needed, e.g, for the contributions quadratic in spin to the post-Newtonian dynamics, see \cite{Porto:Rothstein:2008,Steinhoff:Hergt:Schaefer:2008,Steinhoff:Schaefer:2009,Hergt:Schaefer:2008}. The approach to higher order post-Newtonian spin dynamics in \cite{Tagoshi:etal:2001,Faye:etal:2006,Blanchet:etal:2006,Blanchet:etal:2007} can incorporate quadrupole corrections in a straightforward way. Further, it is possible to derive a canonical formalism from the energy-momentum tensor given in the present paper by the procedure outlined in \cite{Steinhoff:Schaefer:Hergt:2008,Steinhoff:Wang:2009} in certain cases, c.f., \cite{Steinhoff:Hergt:Schaefer:2008}. Canonical methods also proved to be very useful in the post-Minkowskian approximation of single-pole objects, see, e.g., \cite{Ledvinka:Schaefer:Bicak:2008}. Our results may also be used as input for perturbation methods \cite{Mino:etal:1996:1,Tanaka:etal:1996:1,Mino:etal:1997:1,Mino:etal:1997:2,Mino:etal:1997:3} which aim for a description of systems with high mass ratios. Such methods have already been applied to single-pole as well as to pole-dipole objects.  

\subsection{Regularization}\label{open_problems_subsubsec_reg}

A straightforward application of the results in the present paper to self-gravitating compact objects is quite subtle from a mathematical point of view. For short, a strict mathematical definition of the product of distributions does not exist, but would be required due to the non-linearity of Einstein's field equations. Therefore a distributional energy-momentum tensor as a source of the gravitational field makes mathematically no sense in General Relativity. However, this problem can be overcome, as in quantum field theory, by a regularization and renormalization program. In particular, dimensional regularization \cite{tHooft:Veltman:1972,Bollini:Giambiagi:1972} is most useful for theories involving gauge freedoms, like General Relativity. Dimensional regularization has been employed successfully in post-Newtonian calculations \cite{Damour:Jaranowski:Schaefer:2001,Damour:Jaranowski:Schaefer:2008,Blanchet:Damour:Esposito-Farese:2004} to a high order of non-linearity. Notice that all results in the present paper hold for an arbitrary
dimension of spacetime.

\begin{acknowledgments}
This work was supported by the Deutsche Forschungsgemeinschaft (DFG) through the SFB/TR7 ``Gravitational Wave Astronomy'' and GRK 1523. The authors thank G.\ Sch\"afer (Univ.\ Jena) for helpful discussions and comments. DP is greatly indebted to Y.N.\ Obukhov (Moscow State \& Univ.\ Coll.\ London) for many stimulating discussions and constructive criticism. Furthermore, DP would like to thank A.\ Trautman (Warsaw Univ.), W.G.\ Dixon (Univ.\ Cambridge), and W.\ Tulczyjew (INFN Naples) for sharing their insight into gravitational multipole formalisms. Last but not least DP acknowledges the support by B.F.\ Schutz (AEI Golm) as well as the gravitational wave group at the AEI Golm. AEI publication number 2009 - 094.
\end{acknowledgments}

\appendix

\section{From $I$-moments to $J$-moments} \label{Jmoments}

There is a one-to-one transformation from the $I$-moments to the $J$-moments, which are used interchangeably in Dixon's work \cite{Dixon:1974:1,Dixon:1979}. At the quadrupole order it is explicitly given by
\begin{eqnarray}
J^{abcd} &=& I^{[a[cb]d]} := \frac{1}{2} \left( I^{a[c|b|d]} - I^{b[c|a|d]} \right) , \\
I^{abcd} &=& - \frac{4}{3} J^{d(ab)c} = - \frac{4}{3} J^{a(dc)b},
\end{eqnarray}
where $J$ has the following properties:
\begin{eqnarray}
J^{abcd}   &=& J^{[ab][cd]} = J^{cdab}, \\
J^{[abc]d} &=& 0 \quad \Leftrightarrow \quad J^{abcd} + J^{bcad} + J^{cabd} = 0.
\end{eqnarray}
Thus, $J^{abcd}$ has the same (algebraic) symmetries as the Riemann tensor. Equation (\ref{Idecomp}) now becomes
\begin{eqnarray}
J^{abcd} &=& Q^{[a[cb]d]} - 2 u^{[a} Q^{b][cd]} - 2 u^{[c} Q^{d][ab]} \nonumber \\ 
         &&- 3 u^{[a} Q^{b][c} u^{d]}. \label{Jdecomp}
\end{eqnarray}
In terms of $J$ the equations of motion turn into
\begin{eqnarray}
\frac{\delta}{ds} S^{ab} &=& 2 p^{[a} u^{b]} - \frac{4}{3} R^{[a}{}_{cde} J^{b]cde}, \\
\frac{\delta}{ds} p_a    &=& \frac{1}{2} R_{abcd} u^b S^{cd}	+ \frac{1}{6} \nabla_a R_{bcde} J^{bcde},
\end{eqnarray}
and the energy-momentum tensor becomes
\begin{eqnarray}
\widetilde{T}^{ab} &=& \int \left( u^{(a} p^{b)} - \frac{1}{3} R_{cde}{}^{(a} J^{b)edc} \right) \delta_{(4)}	\nonumber \\
&& - \int \nabla_c \left( S^{c(a} u^{b)}\delta_{(4)} \right) 	 \nonumber \\
&& - \frac{2}{3} \int \nabla_d \nabla_c \left( J^{d(ab)c} \delta_{(4)} \right).
\end{eqnarray}

\section{Combined quantities in terms of $\stackrel{2}{o}$ moments} \label{combined_quant_app}

In this appendix we provide a summary of combined quantities at the quadrupole order in terms of the $\stackrel{2}{o}$ moments, as used in the original decomposition of $t^{abcd}$ in (\ref{vel_decomp_quadrupole}).

\begin{eqnarray}
&&I^{dcab} = 2 \left( \stackrel{2}{o}{\!}^{dcab}+ 2 \stackrel{2}{o}{\!}^{dc(a} u^{b)} + 2 \stackrel{2}{o}{\!}^{ab(d} u^{c)}	\right. \nonumber \\ 
&& \left. + \stackrel{2}{o}{\!}^{dc} u^a u^b + \stackrel{2}{o}{\!}^{ab} u^d u^c	- 2 u^{(d} \stackrel{2}{o}{\!}^{c)(a} u^{b)} \right), \label{I_explicit_app} \\ 
&&\stackrel{2}{S}{\!}^{ab} = - 2 \left[ \stackrel{1}{o}{\!}^{[ab]} + \stackrel{1}{o}{\!}^{[a} u^{b]} - 2 \left( \stackrel{2}{o}{\!}^{c[a} u^{b]} + \stackrel{2}{o}{\!}^{c[ab]} \right) \dot{u}_c	\right. \nonumber \\
&& \left. + u_c \stackrel{2}{t}{\!}^{c[a} \dot{u}^{b]} + 2 u_c \frac{ \delta \stackrel{2}{t}{\!}^{[ab]c}}{d s} + 2 u_c u_d \frac{ \delta \stackrel{2}{t}{\!}^{cd[a}}{d s} u^{b]} \right], \label{S2_explicit_app} \\
&&\stackrel{2}{p}{\!}^b = \stackrel{2}{m} u^b + u_a \frac{\delta}{ds} \stackrel{2}{S}{\!}^{ba} + R_{ace}{}^d \left( \frac{4}{3} \stackrel{2}{o}{\!}^{abce} u_d  \right. \nonumber \\ 
&& \left. + 4 \stackrel{2}{o}{\!}^{bae} u^c u_d + \frac{4}{3} \stackrel{2}{o}{\!}^{aec} \rho^b_d + 2 \stackrel{2}{o}{\!}^{ae} u^c \rho^b_d \right), \label{p2_explicit_app} \\
&&\stackrel{2}{m} = \stackrel{0}{t} + \stackrel{2}{S}{\!}^{ab} \dot{u}_a u_b - \frac{2}{3} R_{abc}{}^d \stackrel{2}{o}{\!}^{cba} u_d \nonumber \\
&&	+ u_e u_f \left[ \frac{\delta}{d s} \stackrel{1}{t}{\!}^{ef} - \frac{\delta^2}{d s^2} \stackrel{2}{t}{\!}^{ef} + 2 \frac{\delta}{d s} \left( u_g \frac{\delta }{d s} \stackrel{2}{t}{\!}^{gef} \right) \right. \nonumber \\ 
&&\left. + 2 u^g \stackrel{2}{t}{\!}^{cd(e} R_{gcd}{}^{f)} \right].\label{m2_explicit_app} 
\end{eqnarray}

\section{Conventions \& Symbols}\label{dimension_acronyms_app}

In table \ref{tab_symbols} we provide a list of symbols used throughout the text. Our convention for the signature of spacetime is $-2$. The curvature tensor is defined by (\ref{antisym_cov_deriv}), or equivalently by
\begin{eqnarray}
R_{acd}{}^b:=\Gamma_{dc}{}^b{}_{,a} - \Gamma_{da}{}^b{}_{,c} + \Gamma_{dc}{}^e \Gamma_{ea}{}^b - \Gamma_{da}{}^e \Gamma_{ec}{}^b. 
\end{eqnarray}
The delta function $\delta_{(4)} = \delta_{(4)}( x^a - Y^a(s) )$ is normalized as $\int \delta_{(4)} = 1$.
 
\begin{table}
\caption{\label{tab_symbols}Directory of symbols.} 
\begin{ruledtabular}
\begin{tabular}{ll}
Symbol&Explanation\\
\hline
\multicolumn{2}{l}{{Geometrical quantities}}\\ 
\hline
$g_{a b}$ & Metric\\
$g$ & Determinant of the metric\\
$R_{ijk}{}^l$ & Riemannian curvature\\
$Y^a$ & Worldline within the worldtube of the body \\ 
$u^a$ & Velocity along the worldline $Y^a$ of the body\\
$s$ & Proper time along the worldline\\ 
$\delta_{(4)}$& Four-dimensional delta function\\
&\\
\hline
\multicolumn{2}{l}{{Matter quantities}}\\ 
\hline
$T^{ij}$ &Energy-momentum tensor\\
$t^{ab}, t^{abc}, \dots$ & General multipole moments in the expansion\\
& of the energy momentum density \\ 
$\stackrel{0}{o}{\!}^{a\dots}, \stackrel{0}{t}$ & Parts of the orthogonal decomposition of the \\
& general single-pole moment\\
$\stackrel{1}{o}{\!}^{a\dots}, \stackrel{0}{t}{}^{ab}$ & Parts of the orthogonal decomposition of the \\
& general dipole moment\\
$\stackrel{2}{o}{\!}^{a\dots}, \stackrel{2}{t}{\!}^{a\dots}$ & Parts of the orthogonal decomposition of the \\
& general quadrupole moment\\
$\mathsf{t}^{ab}, \mathsf{t}^{abc}, \dots$ & Canonical multipole moments\\ 
$\stackrel{2}{n}{\!}^{ab}, \stackrel{2}{n}{\!}^{abc}, \dots$ & Parts of the orthogonal decomposition of the \\
& canonical quadrupole moment\\
$\stackrel{0}{m}$ & Mass at the single-pole order \\
$\stackrel{1}{p}{\!}^a$, $\stackrel{1}{S}{\!}^{ab}$& Momentum and spin at the pole-dipole order\\ 
$\stackrel{1}{m}$, $\stackrel{1}{\underline{m}}$ & Masses at the dipole order \\
$\stackrel{1}{E}$ & Conserved quantity at the dipole order (if \\ 
&spacetime has symmetries) \\
$\stackrel{2}{p}{\!}^{a}$, $\stackrel{2}{S}{\!}^{ab}$ & Momentum and spin at the quadrupole order\\
$\stackrel{2}{m}$, $\stackrel{2}{\underline{m}}$ & Masses at the quadrupole order (not conserved \\
& in general)\\
$\stackrel{2}{E}$ & ``Combined'' quantity at the quadrupole order \\ 
&(not conserved in general) \\
$I^{abcd}, J^{abcd}$ & Dixon's quadrupolar moments \\
$m, p^a, S^{ab}$ & Dixon's ``combined'' quantities \\
$Q^{ab}, Q^{abc}, \dots$ & Parts of the orthogonal decomposition of \\
& Dixon's quadrupole moment $I^{abcd}$\\ 
&\\
\hline
\multicolumn{2}{l}{{Operators}}\\ 
\hline
$\nabla_i$, ``$;i$'' &Covariant derivative\\
$\delta/ds$ & Total covariant derivative\\
$\partial_i$, ``$,i$'' &Partial derivative\\
$\rho^a_b$&Spatial projector\\
&\\
\hline
\multicolumn{2}{l}{{Accents}}\\ 
\hline
``$\widetilde{\phantom{pen}}$''&Denotes the density of an object\\
``$\widehat{\phantom{pen}}$''&Denotes the orthogonal projection of an index\\
``$\dot{\phantom{pen}}$''&Denotes the derivative $\delta/ds$\\
\end{tabular}
\end{ruledtabular}
\end{table}

\newpage

\bibliographystyle{unsrtnat}
\bibliography{eom_tulczyjew_bibliography}

\end{document}